\documentclass[%
 reprint,
 amsmath,amssymb,
 aps,showkeys
]{revtex4-2}

\usepackage{graphicx}
\usepackage{dcolumn}
\usepackage{bm}
\usepackage{color}
\usepackage{hyperref}
\usepackage{ulem}

\newcommand{\ii}{{\rm i}}
\newcommand{\dd}{{\rm d}}
\newcommand{\mP}{{\cal P}}
\newcommand{\hw}{{\hat w}}

\newcommand{\hphi}{{\hat\phi}}
\newcommand{\tphi}{{\tilde\phi}}
\newcommand{\hkappa}{{\hat\kappa}}

\newcommand{\cyl}{{\rm cyl.}}

\newcommand{\red}[1]{{#1}}
\def\map{M_{\rm ap} }

\definecolor{Blue}{rgb}{0,0.08,0.65}
\definecolor{Red}{rgb}{0.65,0.08,0.05}
\definecolor{Green}{rgb}{0.15,0.45,0.25}
\definecolor{Purple}{RGB}{153, 51, 153}

\begin{document}

\preprint{APS/123-QED}

\title{Numerical complexity of the joint nulled weak-lensing probability distribution function}

\author{Alexandre Barthelemy}
\altaffiliation{alexandre.barthelemy@iap.fr}
\affiliation{CNRS \& Sorbonne Universit\'e, UMR 7095, Institut d'Astrophysique de Paris, 75014, Paris, France}
\author{Francis Bernardeau}
\affiliation{Institut de Physique Th\'eorique, Universit\'e Paris-Saclay,
CEA, CNRS, UMR 3681, 91191 Gif-sur-Yvette, France}
\author{Sandrine Codis}
\affiliation{AIM, CEA, CNRS, Université Paris-Saclay, Université Paris Diderot, Sorbonne Paris Cité, 91191 Gif-sur-Yvette, France}
\author{Cora Uhlemann}
\affiliation{School of Mathematics, Statistics and Physics, Newcastle University, Herschel Building, NE1 7RU Newcastle-upon-Tyne, United Kingdom}

\date{\today}

\begin{abstract}
In the context of tomographic cosmic shear surveys, there exists a nulling transformation of weak lensing observations (also called BNT transform) that allows us to simplify the correlation structure of tomographic cosmic shear observations, as well as to build observables that depend only on a localised range of redshifts and thus independent from the low-redshift/small-scale modes. This procedure renders possible accurate, and from-first-principles, predictions of the convergence and aperture mass one-point distributions (PDF). We here explore other consequences of this transformation on the (reduced) numerical complexity of the estimation of the joint PDF between nulled bins and demonstrate how to use these results to make theoretical predictions.
\end{abstract}

%Use showkeys class option if keyword
\maketitle

\section{Introduction}

The statistics of weak-lensing fields provide us with a powerful tool for precision cosmology (see e.g. a review in \cite{kilbinger15}) and motivated the build up of new generation large galaxy surveys such as the Legacy Survey of Space and Time (LSST) \citep{LSST} or Euclid \citep{Euclid} which will collect data of unprecedented quality in the coming years. 

Among all the studied statistics of those fields, and especially among probes of their non-Gaussian features, the probability distribution function (PDF) has received some attention. Although some theoretical efforts were engaged long before, some formal reasons can for example be found in \cite{Patton17} who demonstrated that the weak-lensing convergence PDF provides information complementary to the cosmic shear two-point correlation. Note that this complementarity between two-point and non-Gaussian statistics is even more relevant in the presence of 
%systematics such as 
shot and shape noise. Still making use of a Fisher analysis but this time using a "from-first-principles" theoretical model of the convergence PDF, \cite{Boyle2020} also found that it yields tighter constraints for the equation of state of dark energy, the amplitude of fluctuations, the total matter fraction and the sum of neutrino masses, especially when performing a multi-scale analysis and in addition to the two-point correlation function.

As for the theoretical modelling, although there exists numerous works based on numerical simulations or more recently the halo model as in \cite{Thiele20}, we will focus here on the works whose approach could be qualified as "from-first-principles".

Already in \cite{BernardeauValageas} and their following papers, some hierarchical models for the statistics of the matter field were used to construct the PDF of the aperture mass, and these works were later re-interpreted in terms of a large deviation principle and opening the way to extensions to more realistic observables in \cite{paolo}. Note that the general formalism involving large deviation theory was before \cite{paolo} used in the context of the clustering of the matter field itself by the same team, see for example \cite{seminalLDT} or \cite{saddle} for an introduction. Roughly at the same time as \cite{paolo}, the works of \cite{BernardeauValageas} also inspired \cite{FriedrichDES17} for their modelling of the PDF of tangential shear profiles. The most recent works on the PDF of the convergence and the aperture mass within this large deviation framework are \cite{Barthelemy20a} and \cite{Barthelemy21} which take into account the full geometry of the light-cone and probe the validity regime of the formalism with the help of numerical simulations. These papers also make use of a nulling strategy initially developed in \cite{Nulling} and not at all specific to the PDF since directly implemented at the level of the lensing maps.

This nulling strategy, sometimes referred to as the BNT transform, is built on the fact that the lensing window function, that traces the matter density field along the line of sight and gives rise to the observed lensing quantity such as the convergence and the shear within the Born approximation, depends on the
cosmological background evolution
and in particular, does not depend on the dynamical properties of the density fluctuations.
Given this fact and in the context of a tomographic analysis where the lensing fields are observed for sources over a wide range of redshifts, \cite{Nulling} defines a linear transformation -- which thus only depends on the cosmological parameters through two moments only of the lens distance distribution -- that can be applied to the observed maps. It then gives rise to a new set of maps which are effectively the result of a new set of lensing window functions. These \textit{nulled} window functions now vanish over all redshifts comprised between the closest source and the observer, effectively \textit{nulling} the effect of nearby structures. This transformation has multiple consequences: it allows to separate linear or quasi-linear scales from very non-linear scales thus making theoretical modelling easier to control; it also makes nulled lensing maps partially uncorrelated, a property we will use to our advantage here.
These feature are illustrated in Fig.~\ref{fig:pofznum} and further discussed hereafter.

Until now, the nulling strategy has been advocated in the context of multiple nulled maps for the 2-point correlators, angular power spectra or correlation functions, \red{in the context of galaxy and CMB lensing but also line-intensity mapping lensing} \cite{2018PhRvD..98h3514T,2021PhRvD.103d3531T,2021arXiv210609005M}, but not for more complex observables and in particular not for the joint density PDFs.
The goal of this paper is thus now to explore some consequences of the correlation structure of the nulled convergence and aperture mass fields for different source redshifts along the line of sight on the computation of their joint PDF. Note that this makes the derivations we present much more general than this cosmological context since it presents consequences on the computation of the joint PDF of any series of random variables with "blocks" of non-zero inter-correlations, \textit{i.e.} variables that are chain-correlated 2 by 2 or 3 by 3.
Section~\ref{section::convergence} briefly reminds the definition of the convergence and the aperture mass, sections~\ref{section::CGF} and \ref{section::complexity} derive the main results of this paper on the numerical complexity of the joint cumulant generating function (CGF) and PDF, section~\ref{section::implementation} illustrates how one could use these derivations to compute the joint PDF of the nulled convergence field for different sources treating the underlying matter field as in \cite{Barthelemy20a} (large deviation theory), and finally section~\ref{section::conclusion} concludes.

\section{Convergence and Aperture mass}
\label{section::convergence}

The convergence $\kappa$ is traditionally interpreted as a line-of-sight projection of the matter density contrast distribution between the observer and the source. Neglecting lens-lens couplings and within the Born approximation it is written as \citep{kappadef}
\begin{equation}
    \kappa(\bm{\vartheta}) = \int_0^{\chi_s} {\rm d}\chi \, w(\chi,\chi_s) \, \delta(\chi,\mathcal{D}\bm{\vartheta}),
    \label{def-convergence}    
\end{equation}
where $\chi$ is the comoving radial distance -- $\chi_s$ the radial distance of the source -- that depends on the cosmological model, and $\mathcal{D}$ is the comoving angular distance. The lensing kernel $w$ is expressed as
\begin{equation}
\label{eq:weight}
    w(\chi,\chi_s) = \frac{3\,\Omega_m\,H_0^2}{2\,c^2} \, \frac{\mathcal{D}(\chi)\,\mathcal{D}(\chi_s-\chi)}{\mathcal{D}(\chi_s)}\,(1+z(\chi)).
\end{equation}

The aperture mass $M_{\rm ap}$ is defined as a geometrical average of the local convergence within a window of vanishing average
\begin{equation}
    \map(\bm{\vartheta}) = \int {\rm d}^2\bm{\vartheta}' \, U_{\theta}(\vartheta') \, \kappa(\bm{\vartheta}' - \bm{\vartheta})
\end{equation}
with the compensated filter obeying
\begin{equation}
    \int {\rm d}^2\bm{\vartheta}' \, U_{\theta}(\vartheta') = 0.
\end{equation}
The aperture mass can moreover be interestingly expressed as a function of the tangential component $\gamma_t$ of the shear \citep{kkaiser1994,schneider1996}, thus rendering the aperture mass a direct observable up to a reduced shear correction that can also be accounted for as discussed in \cite{Barthelemy21}.

Note that the calculation we present in this paper and especially in section~\ref{section::complexity}, our main result, are very general and in particular their application in the cosmological context does not depend on the choice of a specific filtering scheme for the convergence field. For practical implementation purposes and graphical representation, we will however adopt a simple prescription where the convergence $\kappa_{<\theta}$ is filtered within a top-hat window of angular radius $\theta$ and the associated aperture mass is given by
\begin{equation}
    \map(\bm{\vartheta}) = \kappa_{<\theta_2}(\bm{\vartheta}) - \kappa_{< \theta_1}(\bm{\vartheta}).
    \label{filter_comp}
\end{equation}

In the following we will drop both the mention of $\bm{\vartheta}$, thanks to isotropy, and $< \theta$ to simplify the expressions.

\section{The cumulant generating functions and  the PDFs}
\label{section::CGF}

In order to predict the joint statistics in multiple redshift bins, the  starting point is the expression of the cumulant generating functions of projected densities which will, in our case, correspond to the convergence or the aperture mass.
Following \cite{Barthelemy20a} this cumulant generating function can be shown to take the form,
\begin{equation}
\phi(\{\lambda_{i}\})=\int\dd r\,\phi_{\rm cyl/ slope}\left(\sum_{i}\, \lambda_{i}\,w_{i}(r);r\theta_{0};z(r)\right)\label{phidef}
\end{equation}
where $\theta_{0}$ is the filtering angular size, $w_{i}(r)$ is the line-of-sight profile which leads to the observable $\kappa_{i}$
map, $\phi_{\cyl}\left(\lambda;R;z\right)$ is the cumulant generating function of the density in a cylinder of transverse size $R$, at redshift $z$, written as a function of $\lambda$, and $\phi_{\rm slope}\left(\lambda;R;z\right)$ is the cumulant generating function of the difference between the densities in two concentric cylinders of transverse sizes $R \equiv R_1$ and $R_2 > R_1$ at redshift $z$. The subscripts cyl and slope respectively lead to the convergence and the aperture mass and we write $\phi_{\rm cyl/slope}$ when either or the other can be used. For simplicity, we will present the formalism for the convergence $\kappa$ but the same equations hold for the aperture mass field albeit replacing cyl by slope.

The joint probability density function (PDF) of the maps $\kappa_i$ corresponding to redshift bins indexed by $i$ can then be constructed through an inverse Laplace transformation of the cumulant generating function,
\begin{equation}
\label{eq:Laplace}
\mP\left(\{\kappa_{i}\}\right)=
\int\prod_{i=1}^{n_{t}}\frac{\dd \lambda_{i}}{2 \pi\,\ii}\exp\left[-\sum_{i=1}^{n_{t}}\lambda_{i}\kappa_{i}+\phi(\{\lambda_{i}\})
\right]
\end{equation}
where $n_{t}$ is the total number of source planes and where the integrals run along the imaginary axis. The properties of such a transformation have been described in detail in many papers (see for instance \cite{2013arXiv1311.2724B} and references therein). We simply note  here that marginalizing over one 
variable $\kappa_{p}$ is obtained by setting the corresponding variable $\lambda_{p}$ to 0,
\begin{eqnarray}
&&\int\dd\kappa_{p}\mP\left(\{\kappa_{i}\}\right)=
\int
\frac{\dd \lambda_{1}}{2 \pi\,\ii}
\dots
\frac{\dd \lambda_{p-1}}{2 \pi\,\ii}
\frac{\dd \lambda_{p+1}}{2 \pi\,\ii}
\dots
\frac{\dd \lambda_{n_{t}}}{2 \pi\,\ii}
\times\nonumber\\
&&\exp\left[-\sum_{i=1}^{p-1}\lambda_{i}\kappa_{i}
-\sum_{i=p+1}^{n_{t}}\lambda_{i}\kappa_{i}
+\phi(\{\lambda_{i}\}\vert_{\lambda_{p}=0})
\right].\label{onevarmargin}
\end{eqnarray}
This property will be very useful in the following.

\begin{figure}
   \centering
 \includegraphics[width=\columnwidth]{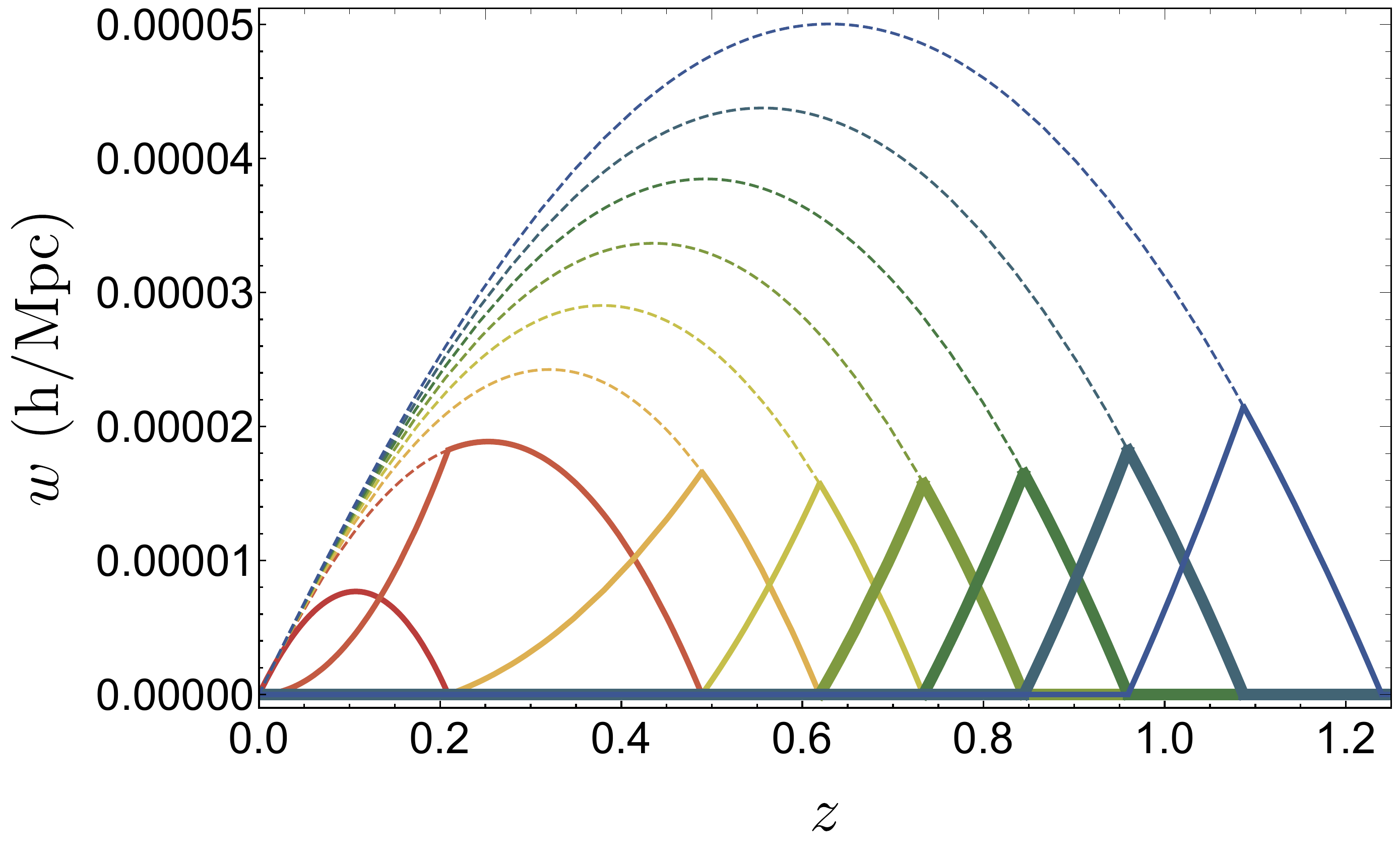}
   \caption{The radial selection  functions, prior to nulling (dashed lines) and after nulling (thick lines). The source redshifts are taken as the mean redshift of the first 8 equally populated redshift bins of the Euclid experiment. The 3 thicker nulled selection function are the ones we use for the practical implementation in section~\ref{section::implementation}.}
   \label{fig:pofznum}
\end{figure}

For simplification we have assumed until now that all sources are located on discrete source planes which makes the selection functions $w_{i}(r)$ specific functions of the radial distance, itself only dependent on the background geometry when working within the Born-approximation. As was previously shown in \cite{Nulling}, it is then possible to define a lower triangular matrix $p_{ij}$ \footnote{to be more precise the only non-zero elements of $p_{ij}$ satisfy $i-2\le j\le i$.} that depends solely on the distance-redshift dependence that linearly transforms the selection functions $w_{i}(r)$ into
\begin{equation}
\hat{w}_{i}(r)=\sum_{j} p_{ij}\, w_{j}(r)
\end{equation}
in such a way that the functions $\hw_{i}(r)$ vanish over all redshifts comprised between the observer and the closest sources. We illustrate this transformation in Fig. \ref{fig:pofznum} where for a set of $\{\kappa_i\}$ and in the case of discrete sources the transformation is possible for all planes except the first 2. The \textit{nulled} selection functions are displayed with solid lines whereas the dashed lines of the same colours represent the corresponding regular selection functions.

Let us now reconsider Equation~(\ref{phidef}) for the variable $\hat{\kappa}_{i}(r)$,
\begin{equation}
\hat\kappa_{i}(r)=\sum_{j} p_{ij}\kappa_{j},
\end{equation}
which only changes $w_{i}(r)$ into $\hw_{i}(r)$ in expression~(\ref{phidef}).
The main integral that  appears in this expression can now be split in $n_t$ different parts - in between the locations of the discrete source planes $\{r_i\}$.
This nulled generating function then reads,
\begin{equation}
\hphi(\{\lambda_{i}\})\!=\!\sum_{n = 1}^{n_t}\int_{r_{n-1}}^{r_n} \!\!\!\!\dd r\,\phi_{\rm cyl/slope}\left(\sum_{i=n}^{n+1}\, \!\!\lambda_{i}\,\hw_{i}(r);r\theta_{0}\right)\label{hphidef}
\end{equation}
where $\lambda_{n_t+1} = 0$ and it is now clear that for each part, only 2 distinct subsequent $\lambda$ variables of each nulled bin appear. In other words the cumulant generating function takes the form
\begin{equation}
\hphi(\{\lambda_{i}\})=\sum_{i=1}^{n_{t}-1}\tphi_{i}(\lambda_{i},\lambda_{i+1})+\tphi_{n_{t}}(\lambda_{n_{t}}).
\label{CGF2}
\end{equation}
Note that the functions $\tphi_{i}$ are not cumulant generating  functions on their own but are closely related to them. By identification and using the fact that $\hphi(\lambda_i)=\hphi(\{\lambda_j=\delta_{ij}\lambda_i\})$,
it can then be shown that we have the following structure,
\begin{eqnarray}
\hphi(\{\lambda_{i}\})&=&\hphi(\lambda_{1},\lambda_{2})-\hphi(\lambda_{2})+\hphi(\lambda_{2},\lambda_{3})-\nonumber\\
&&\dots-\hphi(\lambda_{n_{t}-1})+\hphi(\lambda_{n_{t}-1},\lambda_{n_{t}}),
\label{phisourceplanes}
\end{eqnarray}
where $\hphi$ are now all genuine cumulant generating functions of the corresponding variables.

It thus remains that, with this choice of nulled variables, the joint cumulant generating function has a specific functional form:  it is composed of a sum of functions that depend on 2 variables only, which implies that the full joint PDF can be obtained from bivariate CGFs (and hence PDFs) of neighbouring bins. We will thus now explore the consequences of such a feature on the computations and properties of the joint PDF. 

One can first infer a number of general properties: it is clear that the 2 sets of variables $\left\{\hkappa_{i}\right\}_{i=1,\dots,p-1}$ and $\left\{\hkappa_{i}\right\}_{i=p+1,\dots,n_{t}}$ are correlated only through the variable $\hkappa_{p}$. In other words, once one marginalises over $\hkappa_{p}$, the two sets of variables are independent as there are no common structures that would contribute to both sets. 

At the level of the cumulant generating function, one can notice that we have
\begin{equation}
\hspace{-0.15cm}\hphi\!\left(\{\lambda_{i}\}\vert_{\lambda_{p}=0}\right)\!\!=\! \hphi\!\left(\{\lambda_{i}\}_{i=1,\cdots,p-1}\right)\!+\!
\hphi\!\left(\{\lambda_{i}\}_{i=p+1,\cdots,n_{t}}\right)
\end{equation}
so that the cumulant generating function is split in two separate functions. 

The application of Eq. (\ref{onevarmargin}) then  readily shows that we have,
\begin{eqnarray}
&&\int\dd \hkappa_{p}\,\mP\left(\{\hkappa_{i}\}_{i=1,\dots,n_{t}}\right)=\nonumber \\
&&\hspace{1.5cm}\mP\left(\{\hkappa_{i}\}_{i=1,\dots,p-1}\right)
\,\mP\left(\{\hkappa_{i}\}_{i=p+1,\dots,n_{t}}\right).\label{PFact}
\end{eqnarray}
This independence property for physically separated bins is specific to the nulled variables.

\begin{figure}
   \centering
 \includegraphics[width=\columnwidth]{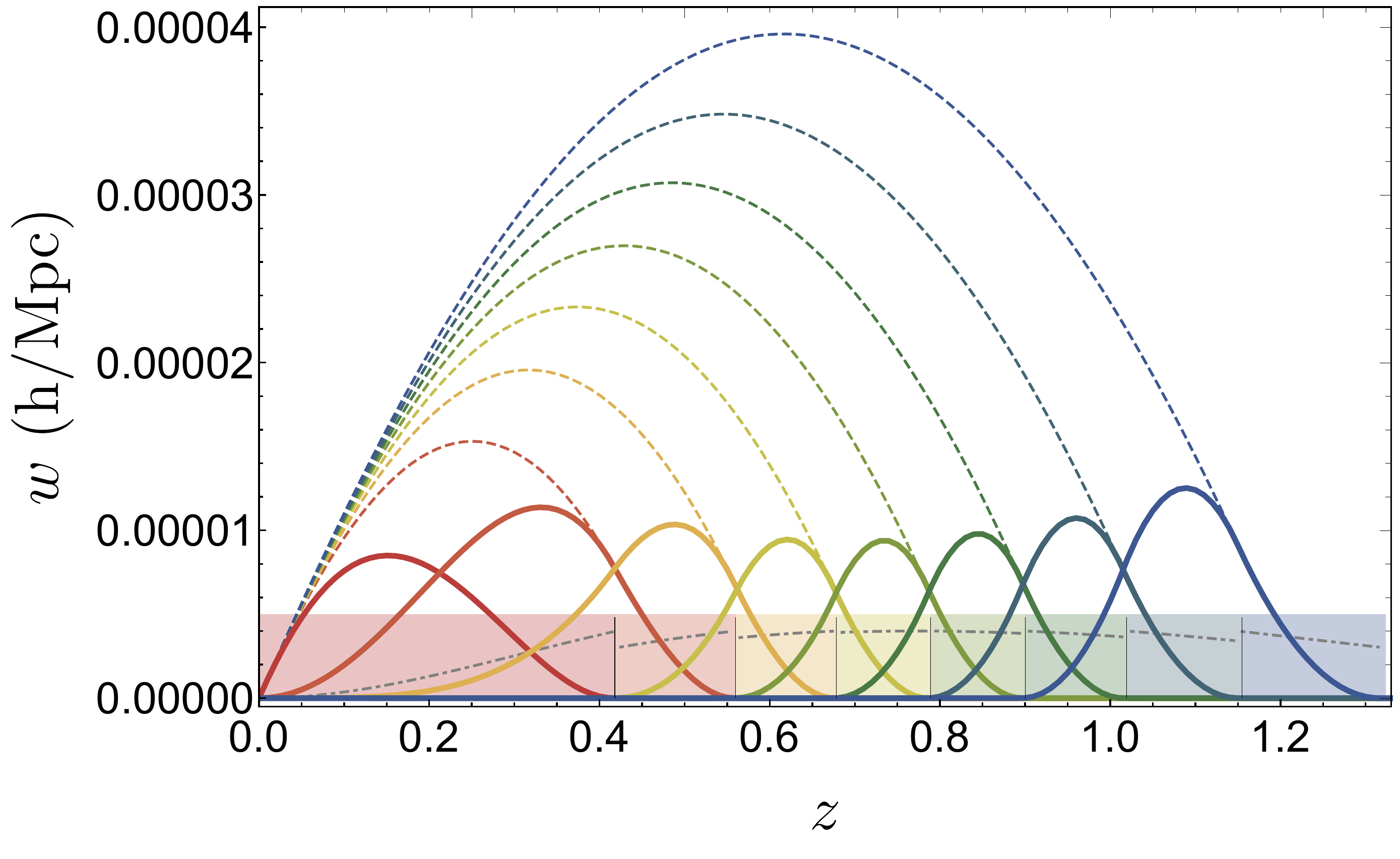}
   \caption{The realistic radial selection functions of the first 8 equally populated redshift bins (shown in coloured rectangles) of the Euclid experiment, prior to nulling (dashed lines) and after nulling (thick lines). The dot-dashed gray lines inside the coloured rectangles show the shape of the normalised distribution of sources inside each bin.
   }
   \label{fig:winulling}
\end{figure} 

Finally, realistic tomographic cases are not made of a collection of discrete source planes but instead source galaxies are split in redshift bins of finite width. The resulting nulling selection functions are then slightly more complex as illustrated in Fig.~\ref{fig:winulling}. One can indeed see here that the nulled profiles overlap with the first and second neighbours - and not only with the first. As a result, the functional form of the cumulant generating function that we obtain is
\begin{eqnarray}
\hphi\left(\{\lambda_{i}\}\right)&=&
\tphi\left(\lambda_{1},\lambda_{2},\lambda_{3}\right)+
\dots+\tphi\left(\lambda_{n_{t}-2},\lambda_{n_{t}-1},\lambda_{n_{t}}\right)\nonumber\\
&&+
\tphi\left(\lambda_{n_{t}-1},\lambda_{n_{t}}\right)+
\tphi\left(\lambda_{n_{t}}\right).\label{CGF3}
\end{eqnarray}
In this case, 
a property similar to Equation~(\ref{PFact}) holds except that an integration over 2 consecutive variables is required.

\subsection*{Impact of shape noise and intrinsic alignment}

The inclusion of a shape noise -- which comes from the variance of the intrinsic ellipticities of galaxies -- to the joint nulled CGF would lead to the addition of a term $\varphi_{\rm noise}(\{\lambda_i\})$ (such that $\hphi_{\rm tot} = \hphi + \varphi$)
that would follow the same functional form (\ref{CGF3}) whether realistic source bins or discrete source planes are considered, with
\begin{equation}
\varphi_{\rm noise}(\{\lambda_i\})=
\sum_{jii'}p_{ij}p_{i'j}\frac{\sigma^2_{\cal S}(j)}{2} \lambda_i\lambda_{i'}
\label{NoiseCGF}
\end{equation}
where $\sigma^2_{\cal S}(j)$ is the amplitude of the shape noise in the original source bin $j$ (and with respect to the shape noise, bins are all independent) assuming it induces a Gaussian noise. This amplitude for the Euclid experiment can for example be found in \cite{2020A&A...636A..95D}. The only non-vanishing terms in (\ref{NoiseCGF}) are those for which there is a $j$ for which $j-2\le i\le j$ and $j-2\le i'\le j$ that implies that $\vert i-i'\vert \le 2$ thus reproducing the form (\ref{CGF3}).

Intrinsic alignments (IA) -- originating from the correlation between the orientation of galaxies and their environment -- represent a contamination for cosmic shear measurements \citep{2015PhR...558....1T}. When the observed shape of a galaxy is decomposed into an intrinsic term and a contribution coming from weak lensing along the line-of-sight, two sorts of correlators arise, the auto-correlation spectra of instrinsic shapes (II in the case of the two-point spectrum) and the cross-correlation spectra between foreground intrinsic shapes and background shear (GI still in the case of the two-point spectrum). There cannot be any contribution coming from foreground shear and background intrinsic shapes, or more generally, the shear-induced ellipticity of a galaxy can only be correlated to objects located within its radial selection function, as those correlations occur on smaller scales (typically below a hundred megaparsecs, therefore not between distant redshift bins). Thus in the case of nulling and similarly to shape noise, intrinsic alignments would also induce correlations that span over 3 nulled bins only, although with a priori a more complex dependence with $\lambda$ since intrinsic alignments may not induce Gaussian noise. Still for the Euclid experiment, a modelling of the impact of intrinsic alignments on the lensing power spectrum and bi-spectra can be found in \cite{2020A&A...636A..95D}, and used to obtain the 2nd and 3rd cumulants in the equivalent of equation~(\ref{NoiseCGF}) for the inclusion of galaxies alignments. However, there is no guarantee that such linear or quasi-linear models are valid in the regime we are probing here and more works is therefore necessary to model this effect properly, including small-scale baryonic effects \citep{2015MNRAS.448.3391C}.

This thus implies that the correlation structure we presented for nulled variables is still valid when such effects are taken into account.  Note however that corrective terms coming from reduced shear, magnification bias or post-born corrections formally break this structure, although weakly, as they change the functional form of equation~(\ref{def-convergence}) and introduce couplings between lenses which are not considered independent anymore. 

Finally note that one obvious but strong limitation of this procedure is the precision to which the redshift of sources inside tomographic bins can be determined (photometric redshift errors). \red{Indeed, in the worst case where the precision of, for example and in the context of a large area survey, photometric redshifts is not very good with a significant number of catastrophic events, the "true" redshift distribution of galaxies inside each redshift bin can become strongly overlapping as is the case for example for the KIDS survey \citep[see figure 2 of][]{2021A&A...645A.104A}. As a result, the nulling procedure cannot be performed perfectly. We expect that for future surveys like Euclid, photometric redshift errors can be reduced as suggested by the treatment of those errors in Euclid collaboration pre-launch papers \citep[see for example equation (46) of][]{2020A&A...636A..95D}. In any case, the quantification of the impact of photometric redshift errors on the nulling procedure is something which can be done rather straightforwardly in the formalism that we present -- computing BNT weights with a given galaxy distribution while having lensing kernel computed from a different one accounting for uncertainties and checking the residual correlations and their impact on say, cosmological parameters estimation. Note also that, as a last-resort backup plan, we could also divide-and-conquer the problem by modelling the photo-z error on the univariate/bivariate/trivariate PDFs and capturing the additional correlations caused by redshift errors in a Gaussian-style copula in the same fashion as what we do in appendix \ref{app::copulas}. Finally, another option could also come from the reliance on some numerical lensing simulation as in \cite{2021MNRAS.506.1623H} to generate systematics-infused control samples and use them to model the impact of photometric redshift and shear calibration uncertainty. This is however an important work on its own and is left for the future as beyond the scope of the results we present here.}

The quantification of all of these effects on the nulling strategy will be performed in future works in preparation.

\section{Numerical complexity for PDF computations}
\label{section::complexity}

Let us now explore the resulting numerical complexity of the computation of the joint PDF of nulled maps, $\hkappa_{i}$.
Note that this is not just a numerical trick to reduce the computing time, this is the simpler form the joint PDF can take using the correlation structures that we have. In particular, equation~(\ref{eq:Laplace}) for nulled bins can as a result be written as some function of joint PDFs between 2 or 3 bins but not in a simpler form than the one we present here, see for example appendix~\ref{app::B}. 

Let us start our analysis in the case of discrete source planes, that is taking advantage of the functional form (\ref{CGF2}) for the cumulant generating function and thus assuming that this generating function can be written as a sum of functions of 2 consecutive variables.

If we assume that the number of operations for a given integral over $\lambda_{i}$ is $N$, then, and in the absence of factorisation properties, the a priori complexity for the computation of the joint $\kappa$-PDF, as given by Equation~\ref{eq:Laplace}, is $N^{n_{t}}$. The purpose of this section is to show how this complexity, and hence the computational time, can be reduced when changing variables from $\kappa$ to $\hkappa$. To reach this goal, let us define the function $\zeta_{2}(n_{t})$,  supposingly an integer, that gives an estimate of the number of operations to be done as $N^{\zeta_{2}(n_{t})}$. Similarly, we also define  a function $\zeta_{3}(n_{t})$ for the  functional form (\ref{CGF3}) of the cumulant generating function corresponding not to successive localised source planes but wide redshift bins.

Let us start with a simple case. For a joint analysis of three source planes, the cumulant generating function takes the form,
\begin{equation}
\hphi(\lambda_{1},\lambda_{2},\lambda_{3})=\hphi(\lambda_{1},\lambda_{2})-\hphi(\lambda_{2})+\hphi(\lambda_{2},\lambda_{3})
\label{eq::18}
\end{equation}
so that we can write,
\begin{equation}
\hspace{-0.2cm}\mP(\hkappa_{1},\hkappa_{2},\hkappa_{3})=\int\frac{\dd\lambda_{2}}{2\pi\,\ii}\,e^{-\lambda_{2}\hkappa_{2}+
\hphi_{\hkappa_{1}}(\lambda_{2})-\hphi(\lambda_{2})+
\hphi_{\hkappa_{3}}(\lambda_{2})}\label{P123}
\end{equation}
where 
\begin{equation}
\label{eq:phikappa}
e^{\hphi_{\kappa_{i}}(\lambda_{j})}=\int\frac{\dd\lambda_{i}}{2\pi\,\ii}\,\exp(-\lambda_{i}\hkappa_{i}+\hphi(\lambda_{i},\lambda_{j})).
\end{equation}
The computation of each function $e^{\hphi_{\kappa_{i}}(\lambda_{j})}$ simply scales like $N^{1}$. This thus implies that the estimated number of operations for the computation of expression (\ref{P123}) is $N\times(N+N)$ which scales like $N^{2}$ so that $\zeta_{2}(3)=2$.

\begin{figure}
   \centering
 \includegraphics[width=\columnwidth]{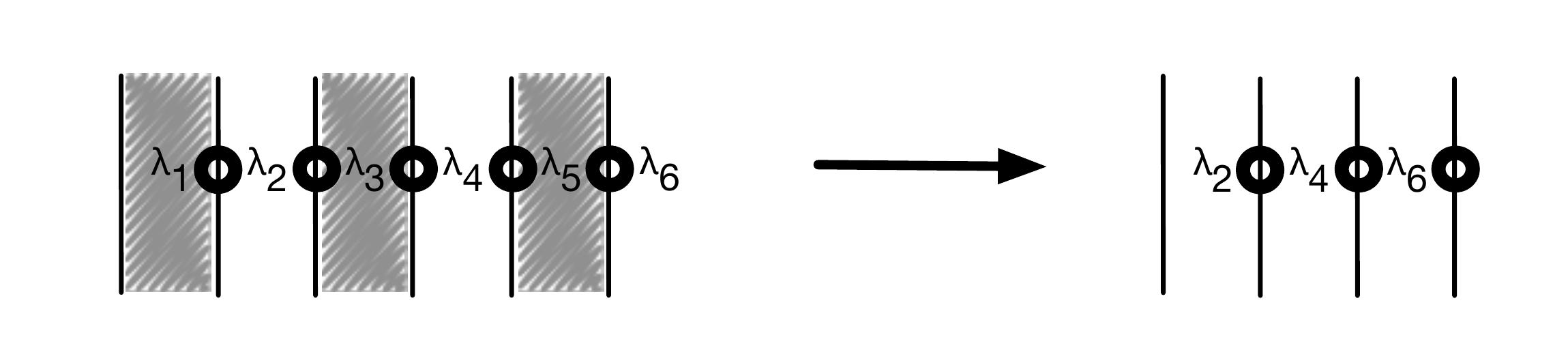}
   \caption{Schematic representation of the decimation procedure for the case of discrete source planes. The generating function is a sum of functions of adjacent variables. Each of these terms is depicted with a dark circle. The integrations over 1 variable out of 2, (grey areas) can then be made independently. The resulting structure is again a sum of functions of adjacent variables and one can thus apply this decimation strategy recursively.}
   \label{fig:JPDFdecimation2}
\end{figure}

Let us now explore the general case and define the quantity 
\begin{eqnarray}
e^{\hphi_{\kappa_{p}}(\lambda_{p-1},\lambda_{p+1})}&=&
\int\frac{\dd\lambda_{p}}{2\pi\,\ii}\,\times\nonumber\\
&&\hspace{-3cm}\exp\left[-\lambda_{p}\hkappa_{p}+\hphi(\lambda_{p-1},\lambda_{p})-\hphi(\lambda_{p})+\hphi(\lambda_{p},\lambda_{p+1})\right].
\end{eqnarray}
The number  of operations for the computation of this function is $N$. 
The idea is now, as exemplified in the simple 3-bin case, to decimate the variables, 1 out of 2, as illustrated on Fig.~\ref{fig:JPDFdecimation2}.

Let us first assume that the number of planes, $n_{t}$, is an odd number, $n_{t}=2m+1$. 
From the structure of the cumulant generating function, and the previous definition, follows one of the main results of this paper 
\begin{widetext}
\begin{equation}
\mP(\{\hkappa_{i}\})=
\int\prod_{i=1}^{m}\frac{\dd\lambda_{2i}}{2\pi\,\ii}\,e^{-\sum_{i=1}^{m}\lambda_{2i}\hkappa_{2i}}\exp\left[\sum_{i=0}^m \hphi_{\kappa_{2i+1}}\!(\lambda_{2i},\lambda_{2(i+1)}) - \hphi(\lambda_{2i})
\right]
\end{equation}
\end{widetext}
with $\lambda_0 = \lambda_{n_t+1} = 0$.
The number of operations required to integrate the odd variables is of the order of $(m+1)N$. What is then remarkable is that the structure of the resulting $m$-inverse Laplace transform is the same as before: the effective cumulant generating function is a sum of function of 2 adjacent variables only. 
As a result the expected number of operations to be performed is $(m+1)N\times N^{\zeta_{2}(m)}$. To a logarithmic correction we will ignore in the following we thus have $\zeta_{2}(2m+1)=1+\zeta_{2}(m)$.
Let us finish this evaluation by noting that there is no gain in reduced numerical complexity going from $2m+1$ to $2m$ source planes. We thus have $\zeta_{2}(2m)=\zeta_{2}(2m+1)=1+\zeta_{2}(m)$. 

Applying this simple rule to the first few integers, we find,
\begin{eqnarray}
\zeta_{2}(1)&=&1\nonumber\\
\zeta_{2}(2)=\zeta_{2}(3)&=&2\nonumber\\
\zeta_{2}(4)=\dots=\zeta_{2}(7)&=&3\nonumber\\
\zeta_{2}(8)=\dots=\zeta_{2}(15)&=&4.
\end{eqnarray}

Finally note again that the reason we need to go through this somewhat complicated numerical integration scheme comes from the fact that there is no simpler form for the joint $\hkappa$-PDF in terms of bivariate PDFs of neighbouring bins as is the case for the CGFs. To explicit this, we show in appendix~\ref{app::B} how the PDF of 3 neighbouring bins is expressed explicitly as a function of bivariate PDFs.

\begin{figure}
   \centering
\includegraphics[width=\columnwidth]{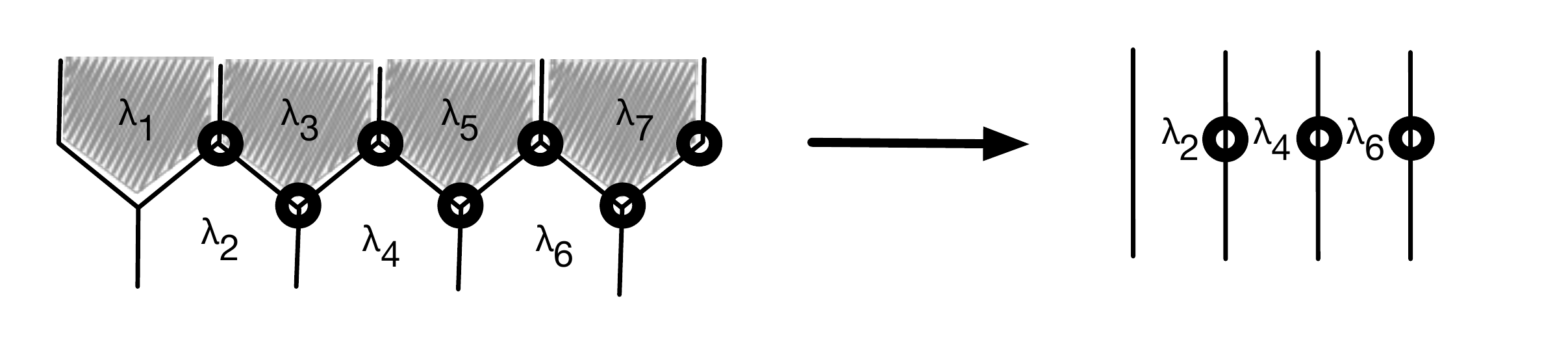}
 \includegraphics[width=\columnwidth]{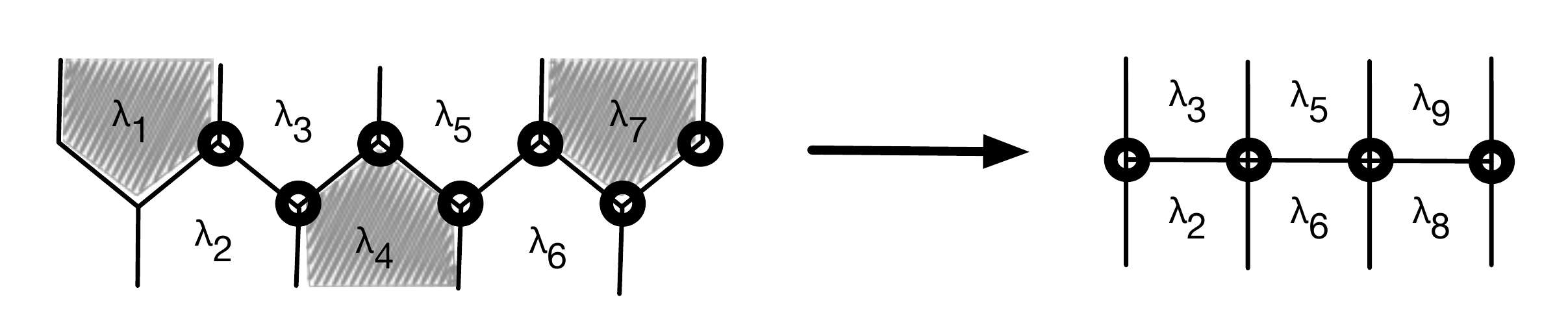}
   \caption{Schematic representation of the decimation procedures for the case of extended source distributions. The generating function is a sum of functions of 3  adjacent variables. Each of these terms are depicted with a dark circle. 
   In the top panel the proposed scheme is to first consider the integration of 1 variable out of 2 (shaded odd variables). These are correlated 2 by 2 and we thus apply the first step depicted in Fig.~\ref{fig:JPDFdecimation2}. We then consider the even variables which are also correlated 2 by 2. This allows to once again perform the first step of Fig.~\ref{fig:JPDFdecimation2} and we then go back to the odd variables and apply this scheme recursively. 
   In the  bottom  panel the procedure is to integrate 1 variable out of 3 where each of those integrations can be performed independently. This leads to an effective cumulant generating function for which variables are correlated via adjacent pairs which we again know how to treat.
   }
   \label{fig:JPDFdecimation3}
\end{figure} 

We now move to the case of equation~ (\ref{CGF3}). We can use the same trick and decimate the variables.  We however identified here 
2 possible strategies which are depicted in Fig.~\ref{fig:JPDFdecimation3}.

The first one is to decimate the variables as before: 1 variables out of 2. The difference here is that this first set of variables are not uncorrelated, they are actually correlated 2 by 2. To be more precise let us again assume that $n_{t}=2m+1$ and the complexity of this operation is thus $N^{\zeta_2(m+1)}$ defined in the previous section. We are then left with $m$ variables that are open again correlated 2 by 2 and therefore have
\begin{equation}
\zeta_{3}(2m+1)=\zeta_2(m+1)+\zeta_2(m)
\end{equation}
and similarly 
\begin{equation}
\zeta_{3}(2m)=2\zeta_2(m).
\end{equation}

The second strategy (the resulting complexity being denoted with a new function $\zeta_3'$) is to decimate 1 variable out of 3, so that the first set of variables are uncorrelated. For this operation, the number of operation scales like $N$. To be more precise let us assume $n_{t}=3m+1$. After decimation we are left with $2m$ variables. These variables then follow a specific structure. They form pairs that are correlated to their nearest neighbours, which is a situation we know  how to handle. Each pair introduce a $N^2$ factor so that we eventually have,
\begin{equation}
\zeta'_{3}(3m+1)=1+2\zeta_{2}(m).
\end{equation}
With such a strategy we have $\zeta'_{3}(3m-1)=\zeta'_{3}(3m)=\zeta_{3}(3m+1)$ (except for $\zeta'_{3}(1)$ and $\zeta'_{3}(2)$).

It turns out that these 2 strategies have  competing performances, as illustrated in Table \ref{zetavalues}.
\begin{table}
\begin{equation}
\begin{array}{|c|ccc|}
\hline
  n_t & \zeta_{2}(n_t) & \zeta_{3}(n_t) & \zeta_{3}'(n_t)\\
  \hline
 1 & 1 & 1 & 1 \\
 2 & 2 & 2 & 2 \\
 3 & 2 & 3 & 3 \\
 4 & 3 & 4 & 3 \\
 5 & 3 & 4 & 5 \\
 6 & 3 & 4 & 5 \\
 7 & 3 & 5 & 5 \\
 8 & 4 & 6 & 5 \\
 9 & 4 & 6 & 5 \\
 10 & 4 & 6 & 5 \\
 11 & 4 & 6 & 7 \\
 12 & 4 & 6 & 7 \\
 13 & 4 & 6 & 7 \\
 14 & 4 & 6 & 7 \\
 15 & 4 & 7 & 7 \\
 \hline
\end{array}\nonumber
\end{equation}
\caption{Numerical complexity $N^{\zeta(n_t)}$ for the calculation of the joint $\kappa$-PDF in case of discrete source planes, $\zeta_{2}(n_t)$, and extended source planes, $\zeta_{3}(n_t)$ and $\zeta_{3}'(n_t)$, as a function of the number of source planes. The latter 2 results correspond to 2 different strategies as described in the main text.
}
\label{zetavalues}
\end{table}

\section{Practical implementation}
\label{section::implementation}

We now move to the actual implementation of the decimation strategy presented in the last section. We will thus start with the computation of $\phi_{\rm cyl}$ and $\phi_{\rm slope}$ and their analytical continuation in the complex plane. The computation of $\hat{\phi}(\{\lambda_i\})$ is then performed on a fixed grid of imaginary values $\{\lambda_i\}$ and we will finally illustrate the procedure described in the previous section with several plots.

\subsection{Computation of $\hat{\phi}(\{\lambda_i\})$}

The computation of this quantity have already been described in several papers, most recently in \cite{Barthelemy20a} for the convergence and \cite{Barthelemy21} for the aperture mass using a from-first-principles formalism inspired by Large deviation theory, though some equations were already known but not interpreted in this framework as in \cite{BernardeauValageas}.

In this context, the joint cumulant generating function of the density filtered in concentric long cylinders of transverse sizes $\{R_i\}$ at redshift $z$ is given by
\begin{equation}
    \phi_{\{\delta_i\}}(\{\lambda_i\}) =  \sum_i \lambda_i\delta_i - \Psi_{\{\delta_i\}}(\{\delta_i\}),
    \label{Legendre}
\end{equation}
where $\{\delta_i\}$ are functions of $\{\lambda_i\}$ through the stationary conditions 
\begin{equation}
   \lambda_k = \frac{\partial \Psi_{\{\delta_i\}}(\{\delta_i\})}{\partial \delta_k}\,,\quad  \forall k.
    \label{stationnary}
\end{equation}
The rate function $\Psi_{\{\delta_i\}}(\{\delta_i\})$ is given by
\begin{equation}
    \Psi_{\{\delta_i\}}(\{\delta_i\})=\frac{1}{2} \sum_{k,j}\Xi_{kj}\mathcal{C}^{-1}(\delta_{k})\mathcal{C}^{-1}(\delta_{j}),
    \label{psicyl}
\end{equation}
where $\Xi_{kj}$ is the inverse of the non-linear covariance matrix between cylinders of transverse radii $\{R_{i} \, (1+\delta_{i})^{1/2}\}$, and the inverse of the cylindrical collapse dynamics $\mathcal{C}$ can be approximately written as 
\begin{equation}
    \mathcal{C}^{-1}(\delta_k) = \nu - \nu (1+\delta_k)^{-1/\nu },\quad \nu = 1.4.
    \label{collapse}
\end{equation}

Then, from equation~(\ref{Legendre}) and the generic properties of cumulant generating functions, one can finally define 
\begin{align}
    \phi_{\rm cyl}(\lambda) &= \phi_{\delta}(\lambda) \quad \text{and} \\
    \phi_{\rm slope}(\lambda) &= \phi_{\delta_1,\delta_2}(-\lambda,\lambda).
\end{align}

The previous set of equations thus allows one to numerically compute $\phi_{\rm cyl}(\lambda)$ and $\phi_{\rm slope}(\lambda)$ for any real value of $\lambda$. Unfortunately, since the computation of the PDF of projected densities requires an integration in the complex plane, we need a prescription for the analytic continuation of $\phi_{\rm cyl/slope}$. As previously explained in \cite{BernardeauValageas} and \cite{Barthelemy21}, a good strategy to keep the analytical properties of the construction~(\ref{Legendre}) consists in fitting an effective $\mathcal{C}$ function rewriting equation~(\ref{Legendre}) for $\phi_{\rm cyl/slope}$ as
\begin{equation}
    \phi_{\rm cyl/slope}(\lambda) = \lambda \mathcal{C}(\tau_{\rm eff}) - \frac{1}{2}\tau_{\rm eff}^2,
\end{equation}
with the effective stationary condition then written as
\begin{equation}
    \lambda = \frac{\rm d}{\rm d \mathcal{C}} \frac{\tau_{\rm eff}^2}{2} = \tau_{\rm eff} \left(\frac{{\rm d}\mathcal{C}(\tau_{\rm eff})}{{\rm d}\tau_{\rm eff}}\right)^{-1}.
    \label{EffectiveStationary}
\end{equation}
Note then that since
\begin{equation}
    \frac{{\rm d}\phi_{\rm cyl/slope}(\lambda)}{{\rm d}\lambda} = \mathcal{C}(\tau_{\rm eff}),
\end{equation}
and hence
\begin{equation}
    \frac{1}{2}\tau_{\rm eff}^2 = \lambda\frac{{\rm d}\phi_{\rm cyl/slope}(\lambda)}{{\rm d}\lambda}-\phi_{\rm cyl/slope}(\lambda),
\end{equation}
one can thus fit both the values of $\tau_{\rm eff}$ and $\mathcal{C}(\tau_{\rm eff})$ from the generating function computed for real values of $\lambda$. An expansion in series of $\mathcal{C}$,
\begin{equation}
    \mathcal{C}(\tau_{\rm eff}) = \sum_{k = 0}^n \frac{\mu_{k}}{k!} \tau_{\rm eff}^k,
\end{equation}
typically stopping at $n=5$, allows to fit the $\mu_k$ coefficients which are closely related to the cumulants of $\phi_{\rm cyl/slope}$ as noted in Appendix D of \cite{Barthelemy21}.

This construction finally enables one to successfully compute $\phi_{\rm cyl/slope}$ and thus the joint cumulant generating function of projected densities $\hat{\phi}(\{\lambda_i\})$ in equation~(\ref{hphidef}) for any tuples of complex $\{\lambda_i\}$.

\subsection{Results}

\begin{figure*}
    \centering
    \includegraphics[width = \columnwidth]{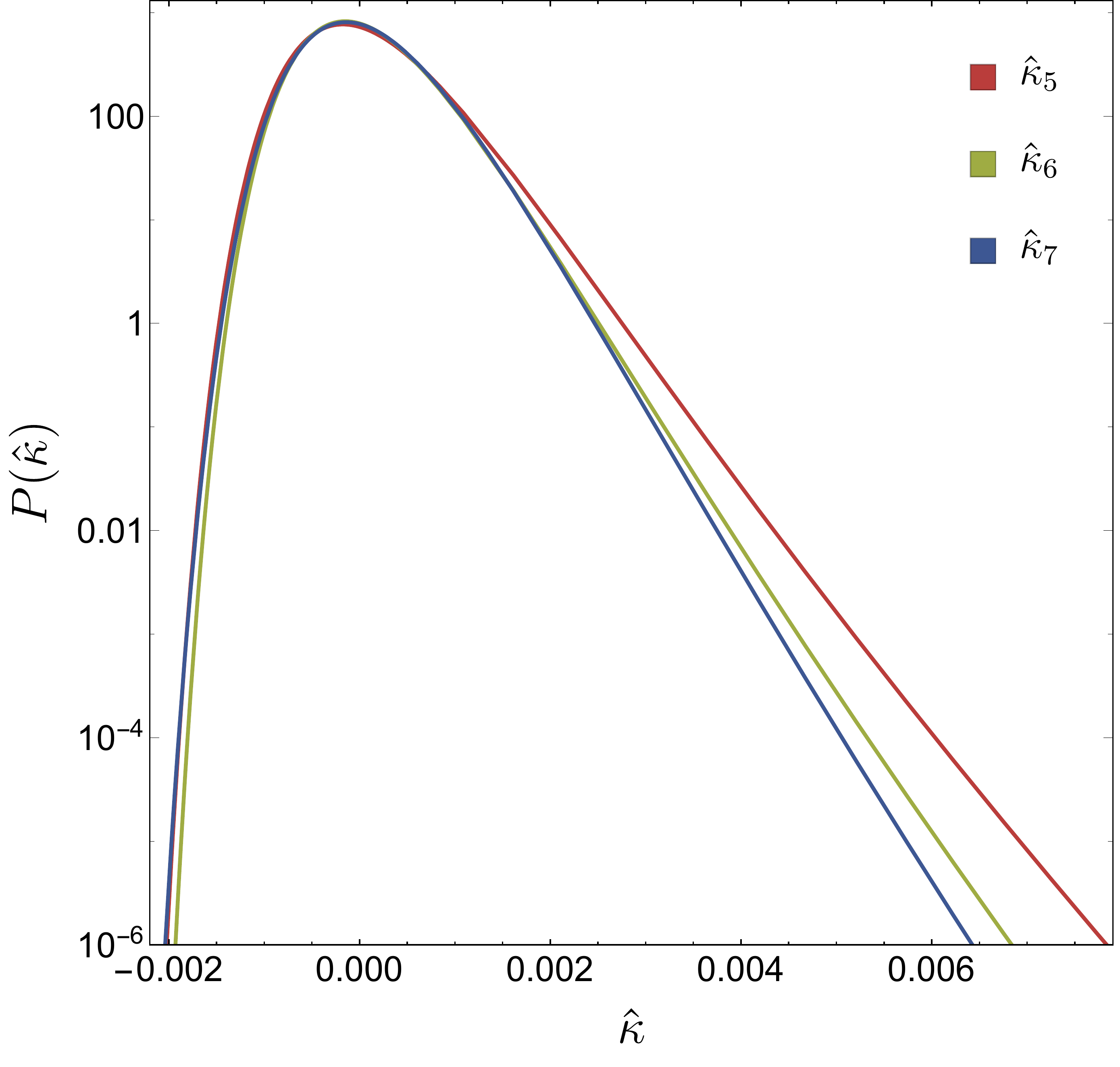}
    \includegraphics[width = \columnwidth]{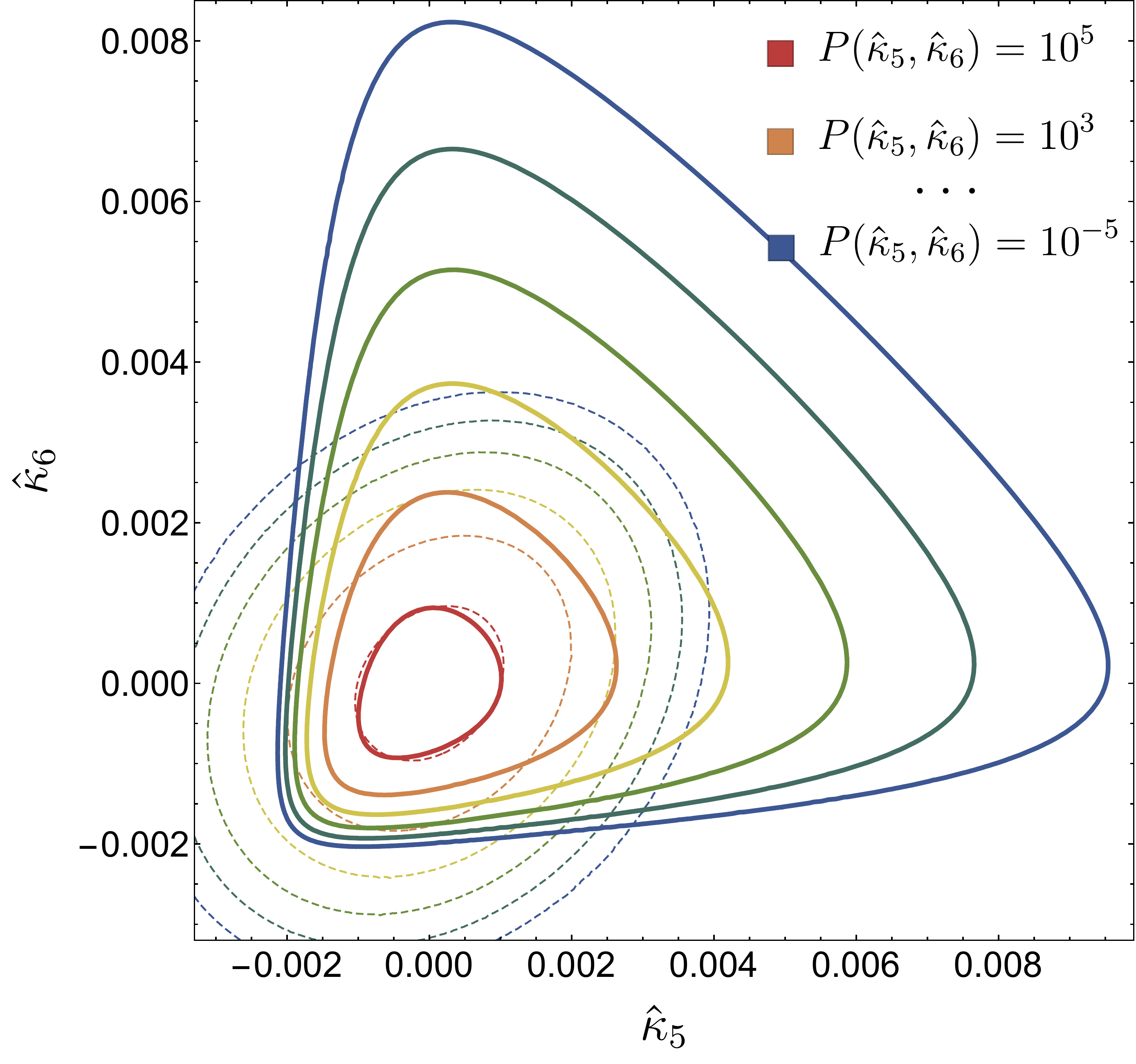}
    \includegraphics[width = \columnwidth]{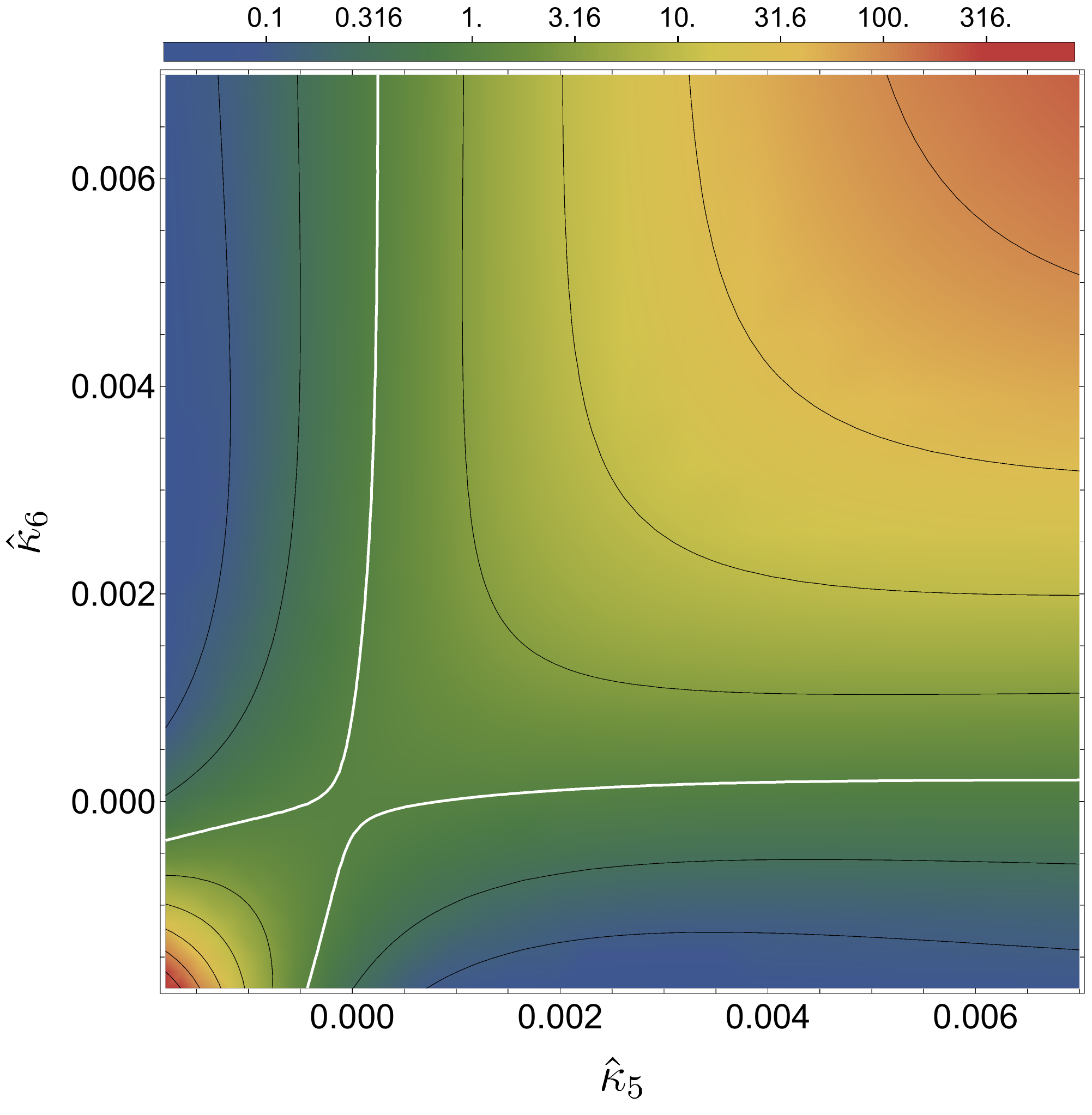}
    \includegraphics[width = \columnwidth]{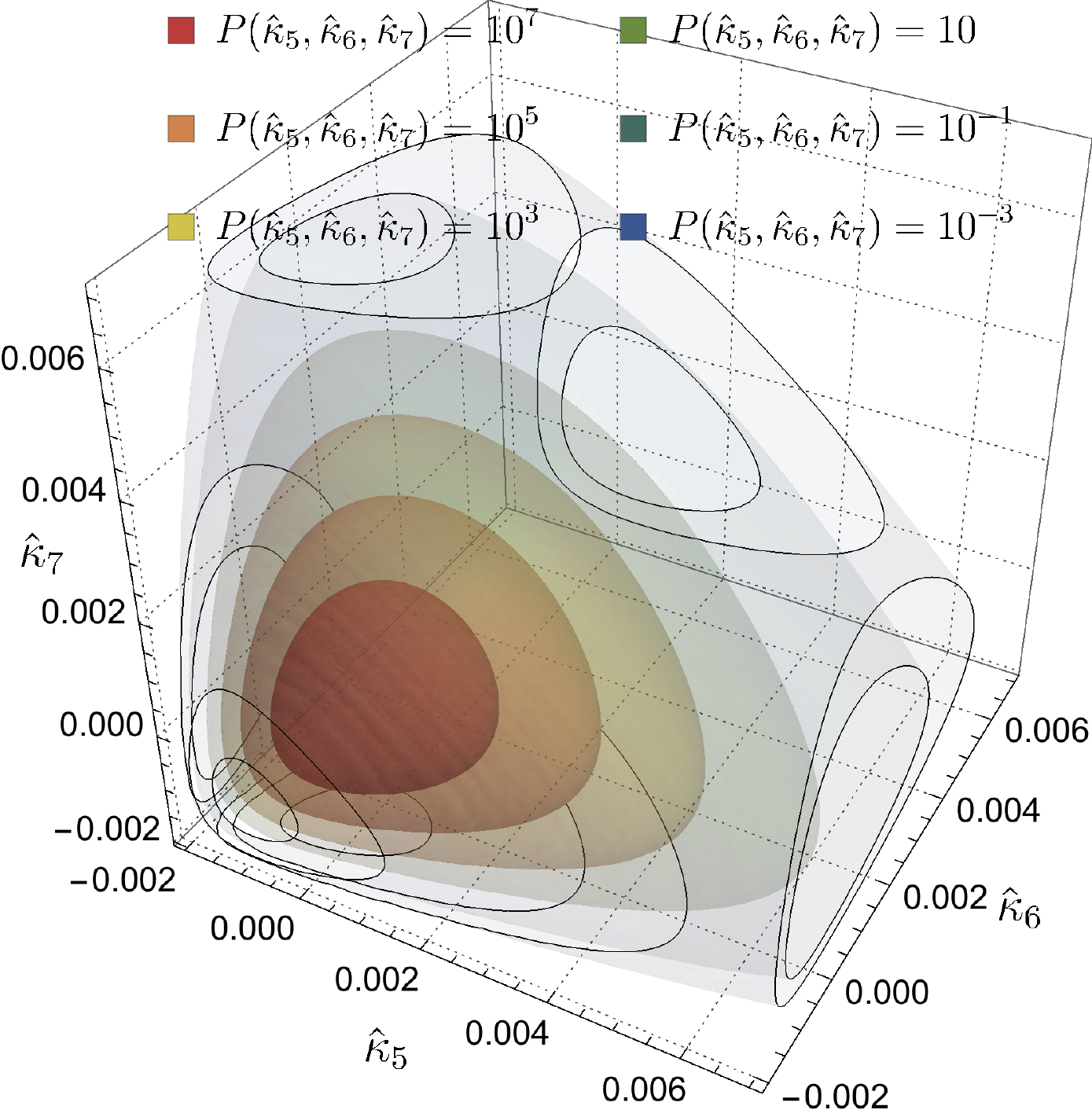}
    \caption{(joint)-PDFs between the 3 nulled bins filtered by a top-hat window function of radius 15 arcmin. Top-left-hand panel: One-point $\hat{\kappa}$-PDF within the 3 nulled bins. Top-right-hand panel: Iso-probability density contours of the joint $\hat{\kappa}$-PDF between the 2 first nulled bins. The nulled fields are filtered by a top-hat window function of radius 15 arcmin. The dashed thin lines correspond to a 2-dimensional normal distribution with the same covariance. Bottom-left-hand-panel: Density plot of $P(\hkappa_5,\hkappa_6)/P(\hkappa_5)/P(\hkappa_6)$ to explicit the shape of the correlation between $\hkappa_5$ and $\hkappa_6$. The black lines correspond to iso-contours of values the ticks of the colour-bar and the white line is for the tick equal to 1. They serve as guide for the eye. Bottom-right-hand panel: Joint $\hat{\kappa}$-PDF between the 3 nulled bins.}
    \label{PDF}
\end{figure*}

For the final part of this paper, we implement the decimation strategy presented in section~\ref{section::complexity} coupled with the computation of $\hat{\phi}(\{\lambda_i\})$ of the last subsection. For simplicity and visualisation purposes we restrict ourselves to the computation of the joint convergence between 3 successive nulled bins taken from Fig.~\ref{fig:pofznum}, the 5th, 6th and 7th, and which are respectively the linear combination of source redshifts located at $z_s = 0.62-0.73-0.85$, $0.73-0.85-0.96$, and $0.85-0.96-1.1$. \red{In practice, the joint PDF is computed from the CGF in a few seconds.}
Thus denoting by respectively $\hkappa_5$, $\hkappa_6$ and $\hkappa_7$ the nulled convergence in each of those bins and noticing that their joint PDF is given by equation~(\ref{P123}) we obtain the results given in Fig.~\ref{PDF} for 1 and 2D marginals and for the full 3D PDF. As expected from the formalism, the exponential decay of the high-$\hkappa$ tails are visible on all these plots and the highly skewed (asymmetric) PDFs hints towards the fact that the information content of the projected densities fields is not entirely probed by its 2-point correlation function which is sufficient for Gaussian random fields only. 

Fig.~\ref{PDF} also displays the shape of the correlation between $\hkappa_5$ and $\hkappa_6$ seen as the residual between the 2D PDF and the product of the two 1D PDFs. This residual would be exactly equal to 1 if the two nulled variables were independent. Note that this modulation of the independent-variables case is quite important and understandable as follows: $\hkappa_5$ and $\hkappa_6$ result from the overlapping contributions of lenses along the line of sight. As a consequence, if $\hkappa_5$ were to take an improbable very high (resp. low) value, then the modulation would have to take that into account by raising the probability of finding a very high (resp. low) value for $\hkappa_6$ compared to the independent-variables case.

\begin{figure*}
    \centering
    \includegraphics[width = \columnwidth]{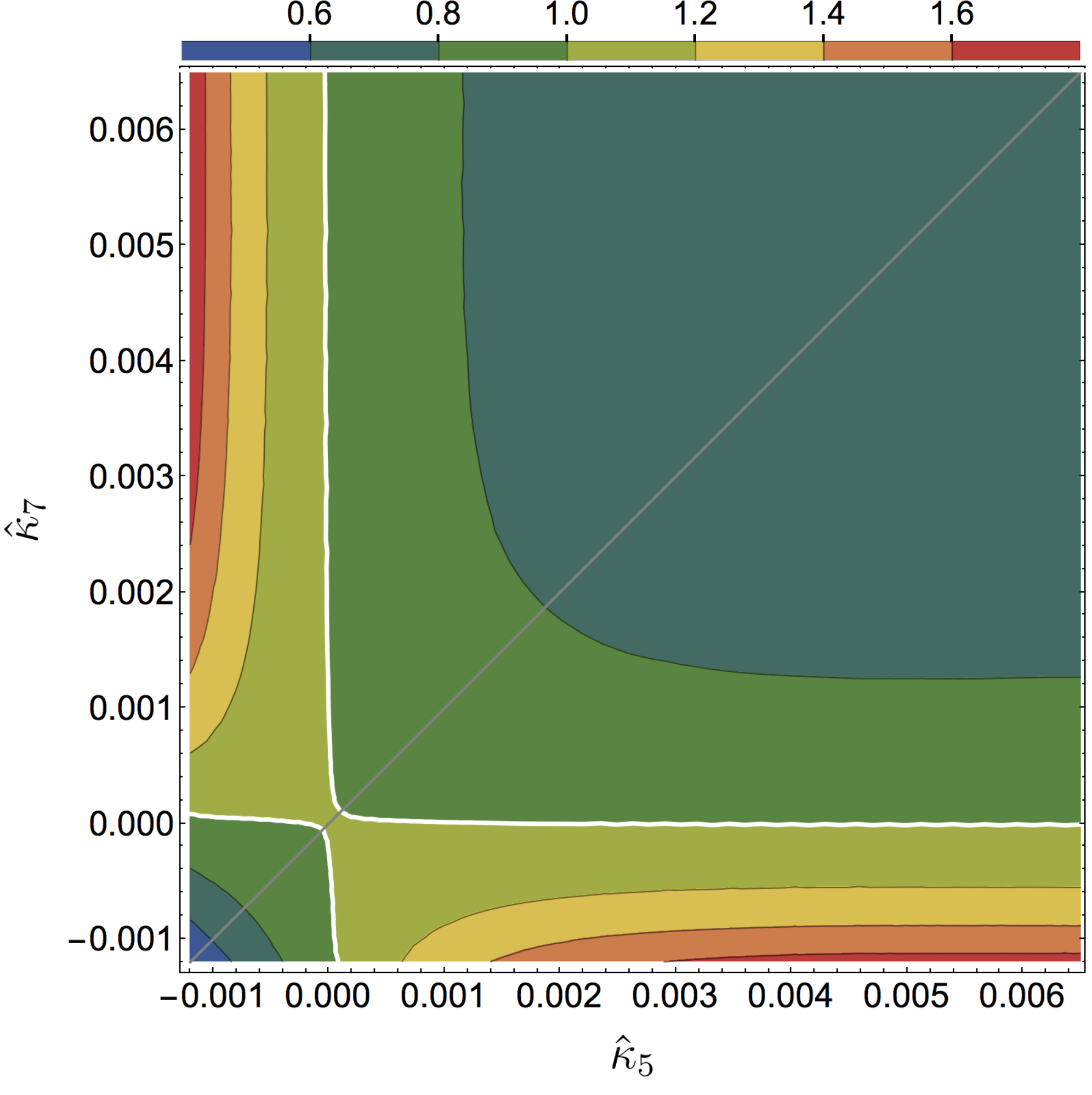}
    \includegraphics[width = \columnwidth]{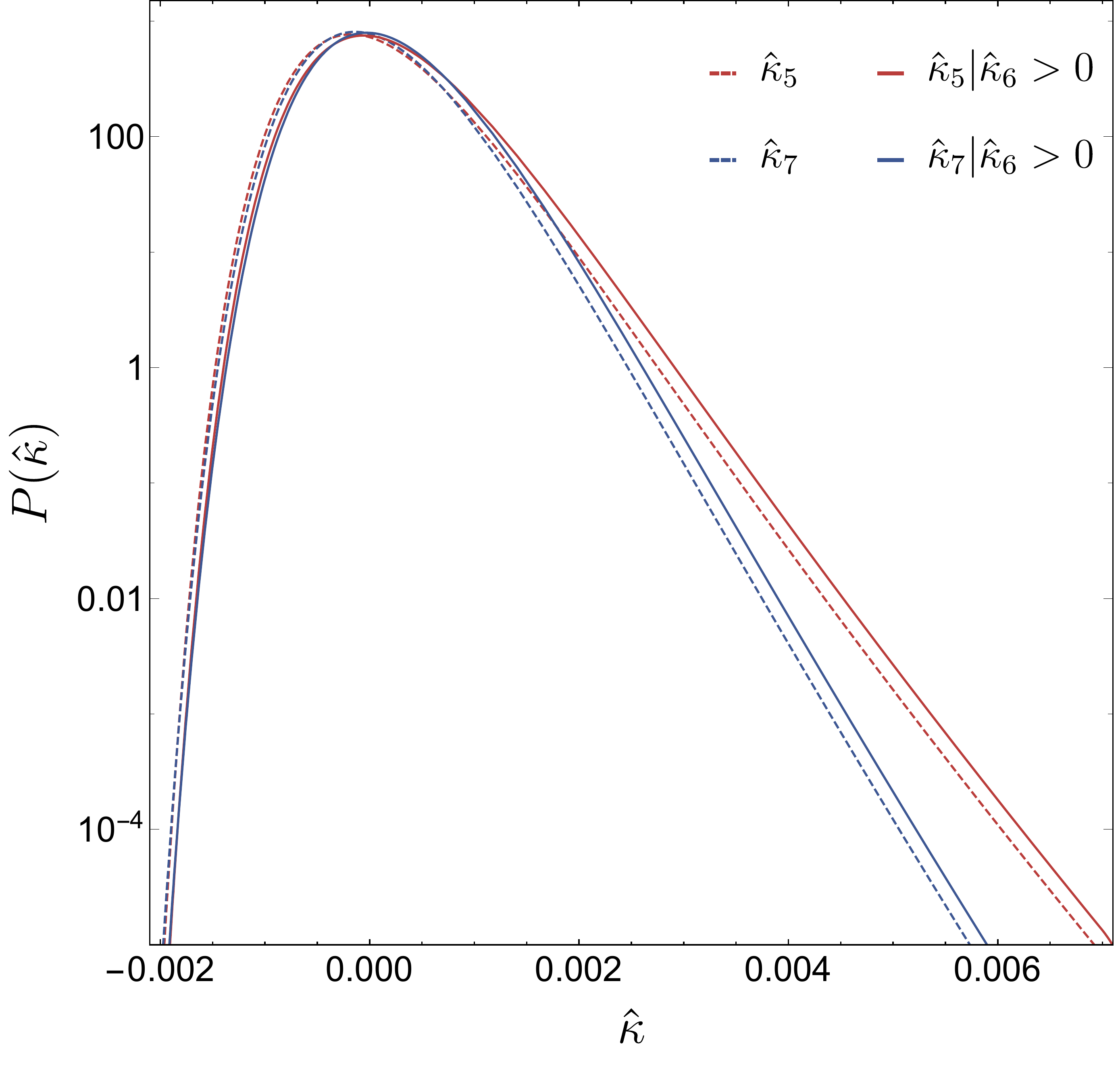}
    \caption{Left-hand panel: Contour plot of $P(\hkappa_5,\hkappa_7 \, | \, \hkappa_6 \! \! > \! \! 0)/P(\hkappa_5 \, | \, \hkappa_6 \! \! > \! \! 0)/P(\hkappa_7 \, | \, \hkappa_6 \! \! > \! \! 0)$ to explicit the shape of the correlation between $\hkappa_5$ and $\hkappa_7$ if a constraint on $\hkappa_6$ is imposed. The white line is for the tick equal to 1. The nulled fields are filtered by a top-hat window function of radius 15 arcmin. Right-hand panel: The solid lines represent $P(\hkappa_5 \, | \, \hkappa_6 \! \! > \! \! 0)$ and $P(\hkappa_7 \, | \, \hkappa_6 \! \! > \! \! 0)$ while the dashed lines show the same PDFs as the top-left panel of Fig.~\ref{PDF}, that is without any constraint. Together the two panels give $P(\hkappa_5,\hkappa_7 \, | \, \hkappa_6 \! \! > \! \! 0)$.}
    \label{residual2}
\end{figure*}

We also show on Fig.~\ref{residual2} the correlation induced between $\hkappa_5$ and $\hkappa_7$ if a constraint on $\hkappa_6$, here $\hkappa_6 > 0$, is imposed. In the absence of constraint, these two variables are independent (since their nulled lensing kernels do not overlap) but the constraint breaks this property. In this case, one needs to compute the full 3-variable PDF before integrating over the constraint. The residual between the 2D constrained PDF and the product of the two 1D constrained PDFs is relatively small compared to the previous case (typically between 50 and 150\%) and points out an over-probability, compared to the independent assumption, for $\hkappa_5$ and $\hkappa_7$ to have opposite rare values while same sign rare events are less likely. Note that since $\hkappa_6$ has its own distribution the constraint $\hkappa_6 > 0$ is more likely to give values of $\hkappa_6$ close to zero and thus the constraint imposes e.g that for a large positive value of $\hkappa_5$, $\hkappa_7$ should compensate in the opposite direction to realise it. Hence the $\hkappa_7$ distribution would be shifted towards more negative values compared to its marginal distribution. For $\hkappa_6 = 0$, $\hkappa_5$ and $\hkappa_7$ would be exactly anti-correlated but this is not exactly the case here as seen from the fact that Fig.~\ref{residual2} is not exactly symetric with respect to the $\hkappa_5 = \hkappa_7$ axis.

\section{Conclusion}
\label{section::conclusion}

This paper explores some properties of the computational complexity of the joint probability distribution function of a set of random variables only correlated by blocks. Building on this specific property we presented different strategies to effectively reduce by several orders of magnitude the number of operations needed to compute this joint probability distribution function. 

In the field of cosmology and in the context of tomographic cosmic shear surveys, such a setup appears when one linearly transforms the set of observed projected densities (convergence, aperture mass) for different source redshifts into their nulled counterparts -- which boils down to rearranging the information present in the survey. As such, the nulled projected densities can be seen as a new basis of pseudo-eigenvectors that render the covariance matrix block-diagonal -- even in the presence of shape noise and intrinsic alignments -- but which also holds the advantage of localising along the line of sight the contribution of structures that make the total signal. To these two main desirable aspects, we now finally add the reduced numerical complexity of the computation of theoretical predictions for the joint PDF of nulled variables. 

Even further reduced numerical complexity could also be achieved using appropriate approximation schemes such as the one we present in appendix~\ref{app::copulas} and that relies on mathematical copulas. This comes at the price of losing the exact result coming from the theory but drastically reduces the computation cost since this scheme only relies in our case on the knowledge of the univariate PDFs and the correlation coefficients, that are strongly linked to the overlap between the lensing kernels.

Regardless of the model -- (semi-)analytical or derived from numerical simulations -- used to infer the statistical properties of the underlying matter field, the nulling formalism ought to be implemented in the context of cosmological analysis of weak lensing surveys. Indeed, while it preserves the information content of the observables, it allows to better identify the origin of this information whether it is in terms of
physical scales, redshift, and combinations of those. In this context, our results show that the actual numerical derivation of joint PDFs is tractable even if one wants to exploit a large number of bins in order to be as precise as possible while preserving the amount of information available. Also note that the method presented in this paper can also be used to obtain the joint PDF of the regular fields via a simple change of variables once the joint PDF of nulled bins has been computed.

For the specific case of non Gaussian observables such as the (joint) PDF of projected densities, it was shown in previous works (see \cite{Barthelemy20a} and \cite{Barthelemy21}) that, for the desired filtering angular scales, shape noise in individual nulled bins makes the detection of non-Gaussian features virtually impossible in a single nulled map. One of the reasons of this failure is that each nulled map collects noise from 3 of the original maps it is built from, see \cite{Barthelemy21} for details.
However, the exploitation of joint PDFs now offers a potential solution to this issue as more useful signal can then be combined together while the shape noise does not increase in the same proportion since it is shared between nulled bins. This thus makes the detection of sought-after non-Gaussian features more likely.

At this stage we can sketch a possible strategy: the key to a successful exploitation of the data is our ability to select the physical scales that are contributing the most to the signal and reject the smallest scales, in practice the nearby lenses, for which we miss reliable modelling and which incidentally are likely to be strongly affected by large super-sample covariance effects. The idea would then be to exploit joined nulled bins, built out of a large set of $n_t$ bins, but a priori ignoring the 2 closest for which the physical scales are still strongly mingled. Hopefully the $n_t-2$ farthest bins would be able to provide us with significant detection of the non-linear couplings in a regime where they are well understood. And if not, it would point to the fact that
the non-Gaussian information that we observe in regular bins is coming from not very well understood nor modelled non-linear scales, which would hence call to extreme caution when using them to extract information. In any case, the nulling formalism and the joint PDF we presented would play a key role in answering these questions. Those studies are however beyond the scope of this paper and will be performed elsewhere.

\section*{Acknowledgements}
This work is partially supported by the SPHERES grant ANR-18-CE31-0009 of the French {\sl Agence Nationale de la Recherche} and by Fondation MERAC.
AB's work is supported by a fellowship from CNES.
This work has made use of the Horizon Cluster hosted by Institut d'Astrophysique de Paris. We thank St\'ephane Rouberol for running smoothly this cluster for us. 
We also thank Oliver Friedrich, Aoife Boyle, Alex Gough and Rapha\"el Gavazzi
for fruitful discussions and comments on the manuscript.
\bibliography{biblio.bib}

\begin{thebibliography}{28}%
\makeatletter
\providecommand \@ifxundefined [1]{%
 \@ifx{#1\undefined}
}%
\providecommand \@ifnum [1]{%
 \ifnum #1\expandafter \@firstoftwo
 \else \expandafter \@secondoftwo
 \fi
}%
\providecommand \@ifx [1]{%
 \ifx #1\expandafter \@firstoftwo
 \else \expandafter \@secondoftwo
 \fi
}%
\providecommand \natexlab [1]{#1}%
\providecommand \enquote  [1]{``#1''}%
\providecommand \bibnamefont  [1]{#1}%
\providecommand \bibfnamefont [1]{#1}%
\providecommand \citenamefont [1]{#1}%
\providecommand \href@noop [0]{\@secondoftwo}%
\providecommand \href [0]{\begingroup \@sanitize@url \@href}%
\providecommand \@href[1]{\@@startlink{#1}\@@href}%
\providecommand \@@href[1]{\endgroup#1\@@endlink}%
\providecommand \@sanitize@url [0]{\catcode `\\12\catcode `\$12\catcode
  `\&12\catcode `\#12\catcode `\^12\catcode `\_12\catcode `\%12\relax}%
\providecommand \@@startlink[1]{}%
\providecommand \@@endlink[0]{}%
\providecommand \url  [0]{\begingroup\@sanitize@url \@url }%
\providecommand \@url [1]{\endgroup\@href {#1}{\urlprefix }}%
\providecommand \urlprefix  [0]{URL }%
\providecommand \Eprint [0]{\href }%
\providecommand \doibase [0]{https://doi.org/}%
\providecommand \selectlanguage [0]{\@gobble}%
\providecommand \bibinfo  [0]{\@secondoftwo}%
\providecommand \bibfield  [0]{\@secondoftwo}%
\providecommand \translation [1]{[#1]}%
\providecommand \BibitemOpen [0]{}%
\providecommand \bibitemStop [0]{}%
\providecommand \bibitemNoStop [0]{.\EOS\space}%
\providecommand \EOS [0]{\spacefactor3000\relax}%
\providecommand \BibitemShut  [1]{\csname bibitem#1\endcsname}%
\let\auto@bib@innerbib\@empty
%</preamble>
\bibitem [{\citenamefont {{Kilbinger}}(2015)}]{kilbinger15}%
  \BibitemOpen
  \bibfield  {author} {\bibinfo {author} {\bibfnamefont {M.}~\bibnamefont
  {{Kilbinger}}},\ }\bibfield  {title} {\bibinfo {title} {{Cosmology with
  cosmic shear observations: a review}},\ }\href
  {https://doi.org/10.1088/0034-4885/78/8/086901} {\bibfield  {journal}
  {\bibinfo  {journal} {Reports on Progress in Physics}\ }\textbf {\bibinfo
  {volume} {78}},\ \bibinfo {eid} {086901} (\bibinfo {year} {2015})},\ \Eprint
  {https://arxiv.org/abs/1411.0115} {arXiv:1411.0115 [astro-ph.CO]}
  \BibitemShut {NoStop}%
\bibitem [{\citenamefont {{Ivezi{\'c}}}\ \emph {et~al.}(2019)\citenamefont
  {{Ivezi{\'c}}}, \citenamefont {{Kahn}}, \citenamefont {{Tyson}},
  \citenamefont {{Abel}}, \citenamefont {{Acosta}}, \citenamefont {{Allsman}},
  \citenamefont {{Alonso}}, \citenamefont {{AlSayyad}}, \citenamefont
  {{Anderson}}, \citenamefont {{Andrew}},\ and\ \citenamefont {et~al.}}]{LSST}%
  \BibitemOpen
  \bibfield  {author} {\bibinfo {author} {\bibfnamefont {{\v{Z}}.}~\bibnamefont
  {{Ivezi{\'c}}}}, \bibinfo {author} {\bibfnamefont {S.~M.}\ \bibnamefont
  {{Kahn}}}, \bibinfo {author} {\bibfnamefont {J.~A.}\ \bibnamefont {{Tyson}}},
  \bibinfo {author} {\bibfnamefont {B.}~\bibnamefont {{Abel}}}, \bibinfo
  {author} {\bibfnamefont {E.}~\bibnamefont {{Acosta}}}, \bibinfo {author}
  {\bibfnamefont {R.}~\bibnamefont {{Allsman}}}, \bibinfo {author}
  {\bibfnamefont {D.}~\bibnamefont {{Alonso}}}, \bibinfo {author}
  {\bibfnamefont {Y.}~\bibnamefont {{AlSayyad}}}, \bibinfo {author}
  {\bibfnamefont {S.~F.}\ \bibnamefont {{Anderson}}}, \bibinfo {author}
  {\bibfnamefont {J.}~\bibnamefont {{Andrew}}},\ and\ \bibinfo {author}
  {\bibnamefont {et~al.}},\ }\bibfield  {title} {\bibinfo {title} {{LSST: From
  Science Drivers to Reference Design and Anticipated Data Products}},\ }\href
  {https://doi.org/10.3847/1538-4357/ab042c} {\bibfield  {journal} {\bibinfo
  {journal} {\apj}\ }\textbf {\bibinfo {volume} {873}},\ \bibinfo {eid} {111}
  (\bibinfo {year} {2019})},\ \Eprint {https://arxiv.org/abs/0805.2366}
  {arXiv:0805.2366 [astro-ph]} \BibitemShut {NoStop}%
\bibitem [{\citenamefont {{Laureijs}}\ \emph {et~al.}(2011)\citenamefont
  {{Laureijs}}, \citenamefont {{Amiaux}}, \citenamefont {{Arduini}},
  \citenamefont {{Augu{\`e}res}}, \citenamefont {{Brinchmann}}, \citenamefont
  {{Cole}}, \citenamefont {{Cropper}}, \citenamefont {{Dabin}}, \citenamefont
  {{Duvet}},\ and\ \citenamefont {{Ealet}}}]{Euclid}%
  \BibitemOpen
  \bibfield  {author} {\bibinfo {author} {\bibfnamefont {R.}~\bibnamefont
  {{Laureijs}}}, \bibinfo {author} {\bibfnamefont {J.}~\bibnamefont
  {{Amiaux}}}, \bibinfo {author} {\bibfnamefont {S.}~\bibnamefont {{Arduini}}},
  \bibinfo {author} {\bibfnamefont {J.~L.}\ \bibnamefont {{Augu{\`e}res}}},
  \bibinfo {author} {\bibfnamefont {J.}~\bibnamefont {{Brinchmann}}}, \bibinfo
  {author} {\bibfnamefont {R.}~\bibnamefont {{Cole}}}, \bibinfo {author}
  {\bibfnamefont {M.}~\bibnamefont {{Cropper}}}, \bibinfo {author}
  {\bibfnamefont {C.}~\bibnamefont {{Dabin}}}, \bibinfo {author} {\bibfnamefont
  {L.}~\bibnamefont {{Duvet}}},\ and\ \bibinfo {author} {\bibfnamefont
  {A.}~\bibnamefont {{Ealet}}},\ }\bibfield  {title} {\bibinfo {title} {{Euclid
  Definition Study Report}},\ }\href@noop {} {\bibfield  {journal} {\bibinfo
  {journal} {arXiv e-prints}\ ,\ \bibinfo {eid} {arXiv:1110.3193}} (\bibinfo
  {year} {2011})},\ \Eprint {https://arxiv.org/abs/1110.3193} {arXiv:1110.3193
  [astro-ph.CO]} \BibitemShut {NoStop}%
\bibitem [{\citenamefont {Patton}\ \emph {et~al.}(2017)\citenamefont {Patton},
  \citenamefont {Blazek}, \citenamefont {Honscheid}, \citenamefont {Huff},
  \citenamefont {Melchior}, \citenamefont {Ross},\ and\ \citenamefont
  {Suchyta}}]{Patton17}%
  \BibitemOpen
  \bibfield  {author} {\bibinfo {author} {\bibfnamefont {K.}~\bibnamefont
  {Patton}}, \bibinfo {author} {\bibfnamefont {J.}~\bibnamefont {Blazek}},
  \bibinfo {author} {\bibfnamefont {K.}~\bibnamefont {Honscheid}}, \bibinfo
  {author} {\bibfnamefont {E.}~\bibnamefont {Huff}}, \bibinfo {author}
  {\bibfnamefont {P.}~\bibnamefont {Melchior}}, \bibinfo {author}
  {\bibfnamefont {A.~J.}\ \bibnamefont {Ross}},\ and\ \bibinfo {author}
  {\bibfnamefont {E.}~\bibnamefont {Suchyta}},\ }\bibfield  {title} {\bibinfo
  {title} {{Cosmological constraints from the convergence 1-point probability
  distribution}},\ }\href {https://doi.org/10.1093/mnras/stx1626} {\bibfield
  {journal} {\bibinfo  {journal} {Monthly Notices of the Royal Astronomical
  Society}\ }\textbf {\bibinfo {volume} {472}},\ \bibinfo {pages} {439}
  (\bibinfo {year} {2017})},\ \Eprint
  {https://arxiv.org/abs/http://oup.prod.sis.lan/mnras/article-pdf/472/1/439/19692830/stx1626.pdf}
  {http://oup.prod.sis.lan/mnras/article-pdf/472/1/439/19692830/stx1626.pdf}
  \BibitemShut {NoStop}%
\bibitem [{\citenamefont {{Boyle}}\ \emph {et~al.}(2020)\citenamefont
  {{Boyle}}, \citenamefont {{Uhlemann}}, \citenamefont {{Barthelemy}},
  \citenamefont {{Friedrich}}, \citenamefont {{Codis}},\ and\ \citenamefont
  {{Bernardeau}}}]{Boyle2020}%
  \BibitemOpen
  \bibfield  {author} {\bibinfo {author} {\bibfnamefont {A.}~\bibnamefont
  {{Boyle}}}, \bibinfo {author} {\bibfnamefont {C.}~\bibnamefont {{Uhlemann}}},
  \bibinfo {author} {\bibfnamefont {A.}~\bibnamefont {{Barthelemy}}}, \bibinfo
  {author} {\bibfnamefont {O.}~\bibnamefont {{Friedrich}}}, \bibinfo {author}
  {\bibfnamefont {S.}~\bibnamefont {{Codis}}},\ and\ \bibinfo {author}
  {\bibfnamefont {F.}~\bibnamefont {{Bernardeau}}},\ }\bibfield  {title}
  {\bibinfo {title} {{Extracting cosmology from the weak lensing convergence
  PDF}},\ }\href@noop {} {\bibfield  {journal} {\bibinfo  {journal} {In prep}\
  } (\bibinfo {year} {2020})}\BibitemShut {NoStop}%
\bibitem [{\citenamefont {{Thiele}}\ \emph {et~al.}(2020)\citenamefont
  {{Thiele}}, \citenamefont {{Hill}},\ and\ \citenamefont
  {{Smith}}}]{Thiele20}%
  \BibitemOpen
  \bibfield  {author} {\bibinfo {author} {\bibfnamefont {L.}~\bibnamefont
  {{Thiele}}}, \bibinfo {author} {\bibfnamefont {J.~C.}\ \bibnamefont
  {{Hill}}},\ and\ \bibinfo {author} {\bibfnamefont {K.~M.}\ \bibnamefont
  {{Smith}}},\ }\bibfield  {title} {\bibinfo {title} {{Accurate analytic model
  for the weak lensing convergence one-point probability distribution function
  and its autocovariance}},\ }\href
  {https://doi.org/10.1103/PhysRevD.102.123545} {\bibfield  {journal} {\bibinfo
   {journal} {\prd}\ }\textbf {\bibinfo {volume} {102}},\ \bibinfo {eid}
  {123545} (\bibinfo {year} {2020})},\ \Eprint
  {https://arxiv.org/abs/2009.06547} {arXiv:2009.06547 [astro-ph.CO]}
  \BibitemShut {NoStop}%
\bibitem [{\citenamefont {{Bernardeau}}\ and\ \citenamefont
  {{Valageas}}(2000)}]{BernardeauValageas}%
  \BibitemOpen
  \bibfield  {author} {\bibinfo {author} {\bibfnamefont {F.}~\bibnamefont
  {{Bernardeau}}}\ and\ \bibinfo {author} {\bibfnamefont {P.}~\bibnamefont
  {{Valageas}}},\ }\bibfield  {title} {\bibinfo {title} {{Construction of the
  one-point PDF of the local aperture mass in weak lensing maps}},\ }\href@noop
  {} {\bibfield  {journal} {\bibinfo  {journal} {Astronomy and Astrophysics}\
  }\textbf {\bibinfo {volume} {364}},\ \bibinfo {pages} {1} (\bibinfo {year}
  {2000})},\ \Eprint {https://arxiv.org/abs/astro-ph/0006270}
  {arXiv:astro-ph/0006270 [astro-ph]} \BibitemShut {NoStop}%
\bibitem [{\citenamefont {{Reimberg}}\ and\ \citenamefont
  {{Bernardeau}}(2018)}]{paolo}%
  \BibitemOpen
  \bibfield  {author} {\bibinfo {author} {\bibfnamefont {P.}~\bibnamefont
  {{Reimberg}}}\ and\ \bibinfo {author} {\bibfnamefont {F.}~\bibnamefont
  {{Bernardeau}}},\ }\bibfield  {title} {\bibinfo {title} {{Large deviation
  principle at work: Computation of the statistical properties of the exact
  one-point aperture mass}},\ }\href
  {https://doi.org/10.1103/PhysRevD.97.023524} {\bibfield  {journal} {\bibinfo
  {journal} {\prd}\ }\textbf {\bibinfo {volume} {97}},\ \bibinfo {eid} {023524}
  (\bibinfo {year} {2018})}\BibitemShut {NoStop}%
\bibitem [{\citenamefont {{Bernardeau}}\ and\ \citenamefont
  {{Reimberg}}(2016)}]{seminalLDT}%
  \BibitemOpen
  \bibfield  {author} {\bibinfo {author} {\bibfnamefont {F.}~\bibnamefont
  {{Bernardeau}}}\ and\ \bibinfo {author} {\bibfnamefont {P.}~\bibnamefont
  {{Reimberg}}},\ }\bibfield  {title} {\bibinfo {title} {{Large deviation
  principle at play in large scale structure cosmology}},\ }\href
  {https://doi.org/10.1103/PhysRevD.94.063520} {\bibfield  {journal} {\bibinfo
  {journal} {\prd}\ }\textbf {\bibinfo {volume} {94}},\ \bibinfo {eid} {063520}
  (\bibinfo {year} {2016})},\ \Eprint {https://arxiv.org/abs/1511.08641}
  {arXiv:1511.08641 [astro-ph.CO]} \BibitemShut {NoStop}%
\bibitem [{\citenamefont {{Uhlemann}}\ \emph {et~al.}(2016)\citenamefont
  {{Uhlemann}}, \citenamefont {{Codis}}, \citenamefont {{Pichon}},
  \citenamefont {{Bernardeau}},\ and\ \citenamefont {{Reimberg}}}]{saddle}%
  \BibitemOpen
  \bibfield  {author} {\bibinfo {author} {\bibfnamefont {C.}~\bibnamefont
  {{Uhlemann}}}, \bibinfo {author} {\bibfnamefont {S.}~\bibnamefont {{Codis}}},
  \bibinfo {author} {\bibfnamefont {C.}~\bibnamefont {{Pichon}}}, \bibinfo
  {author} {\bibfnamefont {F.}~\bibnamefont {{Bernardeau}}},\ and\ \bibinfo
  {author} {\bibfnamefont {P.}~\bibnamefont {{Reimberg}}},\ }\bibfield  {title}
  {\bibinfo {title} {{Back in the saddle: large-deviation statistics of the
  cosmic log-density field}},\ }\href {https://doi.org/10.1093/mnras/stw1074}
  {\bibfield  {journal} {\bibinfo  {journal} {MNRAS}\ }\textbf {\bibinfo
  {volume} {460}},\ \bibinfo {pages} {1529} (\bibinfo {year} {2016})},\ \Eprint
  {https://arxiv.org/abs/1512.05793} {arXiv:1512.05793 [astro-ph.CO]}
  \BibitemShut {NoStop}%
\bibitem [{\citenamefont {{Friedrich}}\ \emph {et~al.}(2018)\citenamefont
  {{Friedrich}}, \citenamefont {{Gruen}}, \citenamefont {{DeRose}},
  \citenamefont {{Kirk}}, \citenamefont {{Krause}}, \citenamefont
  {{McClintock}}, \citenamefont {{Rykoff}}, \citenamefont {{Seitz}},
  \citenamefont {{Wechsler}}, \citenamefont {{Bernstein}}, \citenamefont
  {{Blazek}}, \citenamefont {{Chang}}, \citenamefont {{Hilbert}}, \citenamefont
  {{Jain}}, \citenamefont {{Kovacs}}, \citenamefont {{Lahav}}, \citenamefont
  {{Abdalla}}, \citenamefont {{Allam}}, \citenamefont {{Annis}}, \citenamefont
  {{Bechtol}}, \citenamefont {{Benoit-L{\'e}vy}}, \citenamefont {{Bertin}},
  \citenamefont {{Brooks}}, \citenamefont {{Carnero Rosell}}, \citenamefont
  {{Carrasco Kind}}, \citenamefont {{Carretero}}, \citenamefont {{Cunha}},
  \citenamefont {{D'Andrea}}, \citenamefont {{da Costa}}, \citenamefont
  {{Davis}}, \citenamefont {{Desai}}, \citenamefont {{Diehl}}, \citenamefont
  {{Dietrich}}, \citenamefont {{Drlica-Wagner}}, \citenamefont {{Eifler}},
  \citenamefont {{Fosalba}}, \citenamefont {{Frieman}}, \citenamefont
  {{Garc{\'\i}a-Bellido}}, \citenamefont {{Gaztanaga}}, \citenamefont
  {{Gerdes}}, \citenamefont {{Giannantonio}}, \citenamefont {{Gruendl}},
  \citenamefont {{Gschwend}}, \citenamefont {{Gutierrez}}, \citenamefont
  {{Honscheid}}, \citenamefont {{James}}, \citenamefont {{Jarvis}},
  \citenamefont {{Kuehn}}, \citenamefont {{Kuropatkin}}, \citenamefont
  {{Lima}}, \citenamefont {{March}}, \citenamefont {{Marshall}}, \citenamefont
  {{Melchior}}, \citenamefont {{Menanteau}}, \citenamefont {{Miquel}},
  \citenamefont {{Mohr}}, \citenamefont {{Nord}}, \citenamefont {{Plazas}},
  \citenamefont {{Sanchez}}, \citenamefont {{Scarpine}}, \citenamefont
  {{Schindler}}, \citenamefont {{Schubnell}}, \citenamefont {{Sevilla-Noarbe}},
  \citenamefont {{Sheldon}}, \citenamefont {{Smith}}, \citenamefont
  {{Soares-Santos}}, \citenamefont {{Sobreira}}, \citenamefont {{Suchyta}},
  \citenamefont {{Swanson}}, \citenamefont {{Tarle}}, \citenamefont {{Thomas}},
  \citenamefont {{Troxel}}, \citenamefont {{Vikram}}, \citenamefont
  {{Weller}},\ and\ \citenamefont {{DES Collaboration}}}]{FriedrichDES17}%
  \BibitemOpen
  \bibfield  {author} {\bibinfo {author} {\bibfnamefont {O.}~\bibnamefont
  {{Friedrich}}}, \bibinfo {author} {\bibfnamefont {D.}~\bibnamefont
  {{Gruen}}}, \bibinfo {author} {\bibfnamefont {J.}~\bibnamefont {{DeRose}}},
  \bibinfo {author} {\bibfnamefont {D.}~\bibnamefont {{Kirk}}}, \bibinfo
  {author} {\bibfnamefont {E.}~\bibnamefont {{Krause}}}, \bibinfo {author}
  {\bibfnamefont {T.}~\bibnamefont {{McClintock}}}, \bibinfo {author}
  {\bibfnamefont {E.~S.}\ \bibnamefont {{Rykoff}}}, \bibinfo {author}
  {\bibfnamefont {S.}~\bibnamefont {{Seitz}}}, \bibinfo {author} {\bibfnamefont
  {R.~H.}\ \bibnamefont {{Wechsler}}}, \bibinfo {author} {\bibfnamefont
  {G.~M.}\ \bibnamefont {{Bernstein}}}, \bibinfo {author} {\bibfnamefont
  {J.}~\bibnamefont {{Blazek}}}, \bibinfo {author} {\bibfnamefont
  {C.}~\bibnamefont {{Chang}}}, \bibinfo {author} {\bibfnamefont
  {S.}~\bibnamefont {{Hilbert}}}, \bibinfo {author} {\bibfnamefont
  {B.}~\bibnamefont {{Jain}}}, \bibinfo {author} {\bibfnamefont
  {A.}~\bibnamefont {{Kovacs}}}, \bibinfo {author} {\bibfnamefont
  {O.}~\bibnamefont {{Lahav}}}, \bibinfo {author} {\bibfnamefont {F.~B.}\
  \bibnamefont {{Abdalla}}}, \bibinfo {author} {\bibfnamefont {S.}~\bibnamefont
  {{Allam}}}, \bibinfo {author} {\bibfnamefont {J.}~\bibnamefont {{Annis}}},
  \bibinfo {author} {\bibfnamefont {K.}~\bibnamefont {{Bechtol}}}, \bibinfo
  {author} {\bibfnamefont {A.}~\bibnamefont {{Benoit-L{\'e}vy}}}, \bibinfo
  {author} {\bibfnamefont {E.}~\bibnamefont {{Bertin}}}, \bibinfo {author}
  {\bibfnamefont {D.}~\bibnamefont {{Brooks}}}, \bibinfo {author}
  {\bibfnamefont {A.}~\bibnamefont {{Carnero Rosell}}}, \bibinfo {author}
  {\bibfnamefont {M.}~\bibnamefont {{Carrasco Kind}}}, \bibinfo {author}
  {\bibfnamefont {J.}~\bibnamefont {{Carretero}}}, \bibinfo {author}
  {\bibfnamefont {C.~E.}\ \bibnamefont {{Cunha}}}, \bibinfo {author}
  {\bibfnamefont {C.~B.}\ \bibnamefont {{D'Andrea}}}, \bibinfo {author}
  {\bibfnamefont {L.~N.}\ \bibnamefont {{da Costa}}}, \bibinfo {author}
  {\bibfnamefont {C.}~\bibnamefont {{Davis}}}, \bibinfo {author} {\bibfnamefont
  {S.}~\bibnamefont {{Desai}}}, \bibinfo {author} {\bibfnamefont {H.~T.}\
  \bibnamefont {{Diehl}}}, \bibinfo {author} {\bibfnamefont {J.~P.}\
  \bibnamefont {{Dietrich}}}, \bibinfo {author} {\bibfnamefont
  {A.}~\bibnamefont {{Drlica-Wagner}}}, \bibinfo {author} {\bibfnamefont
  {T.~F.}\ \bibnamefont {{Eifler}}}, \bibinfo {author} {\bibfnamefont
  {P.}~\bibnamefont {{Fosalba}}}, \bibinfo {author} {\bibfnamefont
  {J.}~\bibnamefont {{Frieman}}}, \bibinfo {author} {\bibfnamefont
  {J.}~\bibnamefont {{Garc{\'\i}a-Bellido}}}, \bibinfo {author} {\bibfnamefont
  {E.}~\bibnamefont {{Gaztanaga}}}, \bibinfo {author} {\bibfnamefont {D.~W.}\
  \bibnamefont {{Gerdes}}}, \bibinfo {author} {\bibfnamefont {T.}~\bibnamefont
  {{Giannantonio}}}, \bibinfo {author} {\bibfnamefont {R.~A.}\ \bibnamefont
  {{Gruendl}}}, \bibinfo {author} {\bibfnamefont {J.}~\bibnamefont
  {{Gschwend}}}, \bibinfo {author} {\bibfnamefont {G.}~\bibnamefont
  {{Gutierrez}}}, \bibinfo {author} {\bibfnamefont {K.}~\bibnamefont
  {{Honscheid}}}, \bibinfo {author} {\bibfnamefont {D.~J.}\ \bibnamefont
  {{James}}}, \bibinfo {author} {\bibfnamefont {M.}~\bibnamefont {{Jarvis}}},
  \bibinfo {author} {\bibfnamefont {K.}~\bibnamefont {{Kuehn}}}, \bibinfo
  {author} {\bibfnamefont {N.}~\bibnamefont {{Kuropatkin}}}, \bibinfo {author}
  {\bibfnamefont {M.}~\bibnamefont {{Lima}}}, \bibinfo {author} {\bibfnamefont
  {M.}~\bibnamefont {{March}}}, \bibinfo {author} {\bibfnamefont {J.~L.}\
  \bibnamefont {{Marshall}}}, \bibinfo {author} {\bibfnamefont
  {P.}~\bibnamefont {{Melchior}}}, \bibinfo {author} {\bibfnamefont
  {F.}~\bibnamefont {{Menanteau}}}, \bibinfo {author} {\bibfnamefont
  {R.}~\bibnamefont {{Miquel}}}, \bibinfo {author} {\bibfnamefont {J.~J.}\
  \bibnamefont {{Mohr}}}, \bibinfo {author} {\bibfnamefont {B.}~\bibnamefont
  {{Nord}}}, \bibinfo {author} {\bibfnamefont {A.~A.}\ \bibnamefont
  {{Plazas}}}, \bibinfo {author} {\bibfnamefont {E.}~\bibnamefont {{Sanchez}}},
  \bibinfo {author} {\bibfnamefont {V.}~\bibnamefont {{Scarpine}}}, \bibinfo
  {author} {\bibfnamefont {R.}~\bibnamefont {{Schindler}}}, \bibinfo {author}
  {\bibfnamefont {M.}~\bibnamefont {{Schubnell}}}, \bibinfo {author}
  {\bibfnamefont {I.}~\bibnamefont {{Sevilla-Noarbe}}}, \bibinfo {author}
  {\bibfnamefont {E.}~\bibnamefont {{Sheldon}}}, \bibinfo {author}
  {\bibfnamefont {M.}~\bibnamefont {{Smith}}}, \bibinfo {author} {\bibfnamefont
  {M.}~\bibnamefont {{Soares-Santos}}}, \bibinfo {author} {\bibfnamefont
  {F.}~\bibnamefont {{Sobreira}}}, \bibinfo {author} {\bibfnamefont
  {E.}~\bibnamefont {{Suchyta}}}, \bibinfo {author} {\bibfnamefont {M.~E.~C.}\
  \bibnamefont {{Swanson}}}, \bibinfo {author} {\bibfnamefont {G.}~\bibnamefont
  {{Tarle}}}, \bibinfo {author} {\bibfnamefont {D.}~\bibnamefont {{Thomas}}},
  \bibinfo {author} {\bibfnamefont {M.~A.}\ \bibnamefont {{Troxel}}}, \bibinfo
  {author} {\bibfnamefont {V.}~\bibnamefont {{Vikram}}}, \bibinfo {author}
  {\bibfnamefont {J.}~\bibnamefont {{Weller}}},\ and\ \bibinfo {author}
  {\bibnamefont {{DES Collaboration}}},\ }\bibfield  {title} {\bibinfo {title}
  {{Density split statistics: Joint model of counts and lensing in cells}},\
  }\href {https://doi.org/10.1103/PhysRevD.98.023508} {\bibfield  {journal}
  {\bibinfo  {journal} {\prd}\ }\textbf {\bibinfo {volume} {98}},\ \bibinfo
  {eid} {023508} (\bibinfo {year} {2018})},\ \Eprint
  {https://arxiv.org/abs/1710.05162} {arXiv:1710.05162 [astro-ph.CO]}
  \BibitemShut {NoStop}%
\bibitem [{\citenamefont {{Barthelemy}}\ \emph
  {et~al.}(2020{\natexlab{a}})\citenamefont {{Barthelemy}}, \citenamefont
  {{Codis}}, \citenamefont {{Uhlemann}}, \citenamefont {{Bernardeau}},\ and\
  \citenamefont {{Gavazzi}}}]{Barthelemy20a}%
  \BibitemOpen
  \bibfield  {author} {\bibinfo {author} {\bibfnamefont {A.}~\bibnamefont
  {{Barthelemy}}}, \bibinfo {author} {\bibfnamefont {S.}~\bibnamefont
  {{Codis}}}, \bibinfo {author} {\bibfnamefont {C.}~\bibnamefont {{Uhlemann}}},
  \bibinfo {author} {\bibfnamefont {F.}~\bibnamefont {{Bernardeau}}},\ and\
  \bibinfo {author} {\bibfnamefont {R.}~\bibnamefont {{Gavazzi}}},\ }\bibfield
  {title} {\bibinfo {title} {{A nulling strategy for modelling lensing
  convergence in cones with large deviation theory}},\ }\href
  {https://doi.org/10.1093/mnras/staa053} {\bibfield  {journal} {\bibinfo
  {journal} {MNRAS}\ }\textbf {\bibinfo {volume} {492}},\ \bibinfo {pages}
  {3420} (\bibinfo {year} {2020}{\natexlab{a}})},\ \Eprint
  {https://arxiv.org/abs/1909.02615} {arXiv:1909.02615 [astro-ph.CO]}
  \BibitemShut {NoStop}%
\bibitem [{\citenamefont {{Barthelemy}}\ \emph
  {et~al.}(2020{\natexlab{b}})\citenamefont {{Barthelemy}}, \citenamefont
  {{Codis}},\ and\ \citenamefont {{Bernardeau}}}]{Barthelemy21}%
  \BibitemOpen
  \bibfield  {author} {\bibinfo {author} {\bibfnamefont {A.}~\bibnamefont
  {{Barthelemy}}}, \bibinfo {author} {\bibfnamefont {S.}~\bibnamefont
  {{Codis}}},\ and\ \bibinfo {author} {\bibfnamefont {F.}~\bibnamefont
  {{Bernardeau}}},\ }\bibfield  {title} {\bibinfo {title} {{Probability
  distribution function of the aperture mass field with large deviation
  theory}},\ }\href@noop {} {\bibfield  {journal} {\bibinfo  {journal} {arXiv
  e-prints}\ ,\ \bibinfo {eid} {arXiv:2012.03831}} (\bibinfo {year}
  {2020}{\natexlab{b}})},\ \Eprint {https://arxiv.org/abs/2012.03831}
  {arXiv:2012.03831 [astro-ph.CO]} \BibitemShut {NoStop}%
\bibitem [{\citenamefont {{Bernardeau}}\ \emph {et~al.}(2014)\citenamefont
  {{Bernardeau}}, \citenamefont {{Nishimichi}},\ and\ \citenamefont
  {{Taruya}}}]{Nulling}%
  \BibitemOpen
  \bibfield  {author} {\bibinfo {author} {\bibfnamefont {F.}~\bibnamefont
  {{Bernardeau}}}, \bibinfo {author} {\bibfnamefont {T.}~\bibnamefont
  {{Nishimichi}}},\ and\ \bibinfo {author} {\bibfnamefont {A.}~\bibnamefont
  {{Taruya}}},\ }\bibfield  {title} {\bibinfo {title} {{Cosmic shear full
  nulling: sorting out dynamics, geometry and systematics}},\ }\href
  {https://doi.org/10.1093/mnras/stu1861} {\bibfield  {journal} {\bibinfo
  {journal} {MNRAS}\ }\textbf {\bibinfo {volume} {445}},\ \bibinfo {pages}
  {1526} (\bibinfo {year} {2014})},\ \Eprint {https://arxiv.org/abs/1312.0430}
  {arXiv:1312.0430 [astro-ph.CO]} \BibitemShut {NoStop}%
\bibitem [{\citenamefont {{Taylor}}\ \emph {et~al.}(2018)\citenamefont
  {{Taylor}}, \citenamefont {{Bernardeau}},\ and\ \citenamefont
  {{Kitching}}}]{2018PhRvD..98h3514T}%
  \BibitemOpen
  \bibfield  {author} {\bibinfo {author} {\bibfnamefont {P.~L.}\ \bibnamefont
  {{Taylor}}}, \bibinfo {author} {\bibfnamefont {F.}~\bibnamefont
  {{Bernardeau}}},\ and\ \bibinfo {author} {\bibfnamefont {T.~D.}\ \bibnamefont
  {{Kitching}}},\ }\bibfield  {title} {\bibinfo {title} {{k -cut cosmic shear:
  Tunable power spectrum sensitivity to test gravity}},\ }\href
  {https://doi.org/10.1103/PhysRevD.98.083514} {\bibfield  {journal} {\bibinfo
  {journal} {\prd}\ }\textbf {\bibinfo {volume} {98}},\ \bibinfo {eid} {083514}
  (\bibinfo {year} {2018})},\ \Eprint {https://arxiv.org/abs/1809.03515}
  {arXiv:1809.03515 [astro-ph.CO]} \BibitemShut {NoStop}%
\bibitem [{\citenamefont {{Taylor}}\ \emph {et~al.}(2021)\citenamefont
  {{Taylor}}, \citenamefont {{Bernardeau}},\ and\ \citenamefont
  {{Huff}}}]{2021PhRvD.103d3531T}%
  \BibitemOpen
  \bibfield  {author} {\bibinfo {author} {\bibfnamefont {P.~L.}\ \bibnamefont
  {{Taylor}}}, \bibinfo {author} {\bibfnamefont {F.}~\bibnamefont
  {{Bernardeau}}},\ and\ \bibinfo {author} {\bibfnamefont {E.}~\bibnamefont
  {{Huff}}},\ }\bibfield  {title} {\bibinfo {title} {{x -cut Cosmic shear:
  Optimally removing sensitivity to baryonic and nonlinear physics with an
  application to the Dark Energy Survey year 1 shear data}},\ }\href
  {https://doi.org/10.1103/PhysRevD.103.043531} {\bibfield  {journal} {\bibinfo
   {journal} {\prd}\ }\textbf {\bibinfo {volume} {103}},\ \bibinfo {eid}
  {043531} (\bibinfo {year} {2021})},\ \Eprint
  {https://arxiv.org/abs/2007.00675} {arXiv:2007.00675 [astro-ph.CO]}
  \BibitemShut {NoStop}%
\bibitem [{\citenamefont {{Maniyar}}\ \emph {et~al.}(2021)\citenamefont
  {{Maniyar}}, \citenamefont {{Schaan}},\ and\ \citenamefont
  {{Pullen}}}]{2021arXiv210609005M}%
  \BibitemOpen
  \bibfield  {author} {\bibinfo {author} {\bibfnamefont {A.~S.}\ \bibnamefont
  {{Maniyar}}}, \bibinfo {author} {\bibfnamefont {E.}~\bibnamefont
  {{Schaan}}},\ and\ \bibinfo {author} {\bibfnamefont {A.~R.}\ \bibnamefont
  {{Pullen}}},\ }\bibfield  {title} {\bibinfo {title} {{A new probe of the
  high-redshift Universe: nulling CMB lensing with interloper-free ``LIM-pair''
  lensing}},\ }\href@noop {} {\bibfield  {journal} {\bibinfo  {journal} {arXiv
  e-prints}\ ,\ \bibinfo {eid} {arXiv:2106.09005}} (\bibinfo {year} {2021})},\
  \Eprint {https://arxiv.org/abs/2106.09005} {arXiv:2106.09005 [astro-ph.CO]}
  \BibitemShut {NoStop}%
\bibitem [{\citenamefont {Mellier}(1999)}]{kappadef}%
  \BibitemOpen
  \bibfield  {author} {\bibinfo {author} {\bibfnamefont {Y.}~\bibnamefont
  {Mellier}},\ }\bibfield  {title} {\bibinfo {title} {Probing the universe with
  weak lensing},\ }\href {https://doi.org/10.1146/annurev.astro.37.1.127}
  {\bibfield  {journal} {\bibinfo  {journal} {Annual Review of Astronomy and
  Astrophysics}\ }\textbf {\bibinfo {volume} {37}},\ \bibinfo {pages} {127}
  (\bibinfo {year} {1999})},\ \Eprint
  {https://arxiv.org/abs/https://doi.org/10.1146/annurev.astro.37.1.127}
  {https://doi.org/10.1146/annurev.astro.37.1.127} \BibitemShut {NoStop}%
\bibitem [{\citenamefont {{Kaiser}}(1995)}]{kkaiser1994}%
  \BibitemOpen
  \bibfield  {author} {\bibinfo {author} {\bibfnamefont {N.}~\bibnamefont
  {{Kaiser}}},\ }\bibfield  {title} {\bibinfo {title} {{Nonlinear Cluster Lens
  Reconstruction}},\ }\href {https://doi.org/10.1086/187730} {\bibfield
  {journal} {\bibinfo  {journal} {Astrophysical Journal, Letters}\ }\textbf
  {\bibinfo {volume} {439}},\ \bibinfo {pages} {L1} (\bibinfo {year} {1995})},\
  \Eprint {https://arxiv.org/abs/astro-ph/9408092} {arXiv:astro-ph/9408092
  [astro-ph]} \BibitemShut {NoStop}%
\bibitem [{\citenamefont {{Schneider}}(1996)}]{schneider1996}%
  \BibitemOpen
  \bibfield  {author} {\bibinfo {author} {\bibfnamefont {P.}~\bibnamefont
  {{Schneider}}},\ }\bibfield  {title} {\bibinfo {title} {{Detection of (dark)
  matter concentrations via weak gravitational lensing}},\ }\href
  {https://doi.org/10.1093/mnras/283.3.837} {\bibfield  {journal} {\bibinfo
  {journal} {MNRAS}\ }\textbf {\bibinfo {volume} {283}},\ \bibinfo {pages}
  {837} (\bibinfo {year} {1996})},\ \Eprint
  {https://arxiv.org/abs/astro-ph/9601039} {arXiv:astro-ph/9601039 [astro-ph]}
  \BibitemShut {NoStop}%
\bibitem [{\citenamefont {{Bernardeau}}(2013)}]{2013arXiv1311.2724B}%
  \BibitemOpen
  \bibfield  {author} {\bibinfo {author} {\bibfnamefont {F.}~\bibnamefont
  {{Bernardeau}}},\ }\bibfield  {title} {\bibinfo {title} {{The evolution of
  the large-scale structure of the universe: beyond the linear regime}},\
  }\href@noop {} {\bibfield  {journal} {\bibinfo  {journal} {arXiv e-prints}\
  ,\ \bibinfo {eid} {arXiv:1311.2724}} (\bibinfo {year} {2013})},\ \Eprint
  {https://arxiv.org/abs/1311.2724} {arXiv:1311.2724 [astro-ph.CO]}
  \BibitemShut {NoStop}%
\bibitem [{Note1()}]{Note1}%
  \BibitemOpen
  \bibinfo {note} {To be more precise the only non-zero elements of $p_{ij}$
  satisfy $i-2\le j\le i$.}\BibitemShut {Stop}%
\bibitem [{\citenamefont {{Deshpande}}\ \emph {et~al.}(2020)\citenamefont
  {{Deshpande}}, \citenamefont {{Kitching}}, \citenamefont {{Cardone}},
  \citenamefont {{Taylor}}, \citenamefont {{Casas}}, \citenamefont {{Camera}},
  \citenamefont {{Carbone}}, \citenamefont {{Kilbinger}}, \citenamefont
  {{Pettorino}}, \citenamefont {{Sakr}}, \citenamefont {{Sapone}},
  \citenamefont {{Tutusaus}}, \citenamefont {{Auricchio}}, \citenamefont
  {{Bodendorf}}, \citenamefont {{Bonino}}, \citenamefont {{Brescia}},
  \citenamefont {{Capobianco}}, \citenamefont {{Carretero}}, \citenamefont
  {{Castellano}}, \citenamefont {{Cavuoti}}, \citenamefont {{Cledassou}},
  \citenamefont {{Congedo}}, \citenamefont {{Conversi}}, \citenamefont
  {{Corcione}}, \citenamefont {{Cropper}}, \citenamefont {{Dubath}},
  \citenamefont {{Dusini}}, \citenamefont {{Fabbian}}, \citenamefont
  {{Franceschi}}, \citenamefont {{Fumana}}, \citenamefont {{Garilli}},
  \citenamefont {{Grupp}}, \citenamefont {{Hoekstra}}, \citenamefont
  {{Hormuth}}, \citenamefont {{Israel}}, \citenamefont {{Jahnke}},
  \citenamefont {{Kermiche}}, \citenamefont {{Kubik}}, \citenamefont {{Kunz}},
  \citenamefont {{Lacasa}}, \citenamefont {{Ligori}}, \citenamefont {{Lilje}},
  \citenamefont {{Lloro}}, \citenamefont {{Maiorano}}, \citenamefont
  {{Marggraf}}, \citenamefont {{Massey}}, \citenamefont {{Mei}}, \citenamefont
  {{Meneghetti}}, \citenamefont {{Meylan}}, \citenamefont {{Moscardini}},
  \citenamefont {{Padilla}}, \citenamefont {{Paltani}}, \citenamefont
  {{Pasian}}, \citenamefont {{Pires}}, \citenamefont {{Polenta}}, \citenamefont
  {{Poncet}}, \citenamefont {{Raison}}, \citenamefont {{Rhodes}}, \citenamefont
  {{Roncarelli}}, \citenamefont {{Saglia}}, \citenamefont {{Schneider}},
  \citenamefont {{Secroun}}, \citenamefont {{Serrano}}, \citenamefont
  {{Sirri}}, \citenamefont {{Starck}}, \citenamefont {{Sureau}}, \citenamefont
  {{Taylor}}, \citenamefont {{Tereno}}, \citenamefont {{Toledo-Moreo}},
  \citenamefont {{Valenziano}}, \citenamefont {{Wang}},\ and\ \citenamefont
  {{Zoubian}}}]{2020A&A...636A..95D}%
  \BibitemOpen
  \bibfield  {author} {\bibinfo {author} {\bibfnamefont {A.~C.}\ \bibnamefont
  {{Deshpande}}}, \bibinfo {author} {\bibfnamefont {T.~D.}\ \bibnamefont
  {{Kitching}}}, \bibinfo {author} {\bibfnamefont {V.~F.}\ \bibnamefont
  {{Cardone}}}, \bibinfo {author} {\bibfnamefont {P.~L.}\ \bibnamefont
  {{Taylor}}}, \bibinfo {author} {\bibfnamefont {S.}~\bibnamefont {{Casas}}},
  \bibinfo {author} {\bibfnamefont {S.}~\bibnamefont {{Camera}}}, \bibinfo
  {author} {\bibfnamefont {C.}~\bibnamefont {{Carbone}}}, \bibinfo {author}
  {\bibfnamefont {M.}~\bibnamefont {{Kilbinger}}}, \bibinfo {author}
  {\bibfnamefont {V.}~\bibnamefont {{Pettorino}}}, \bibinfo {author}
  {\bibfnamefont {Z.}~\bibnamefont {{Sakr}}}, \bibinfo {author} {\bibfnamefont
  {D.}~\bibnamefont {{Sapone}}}, \bibinfo {author} {\bibfnamefont
  {I.}~\bibnamefont {{Tutusaus}}}, \bibinfo {author} {\bibfnamefont
  {N.}~\bibnamefont {{Auricchio}}}, \bibinfo {author} {\bibfnamefont
  {C.}~\bibnamefont {{Bodendorf}}}, \bibinfo {author} {\bibfnamefont
  {D.}~\bibnamefont {{Bonino}}}, \bibinfo {author} {\bibfnamefont
  {M.}~\bibnamefont {{Brescia}}}, \bibinfo {author} {\bibfnamefont
  {V.}~\bibnamefont {{Capobianco}}}, \bibinfo {author} {\bibfnamefont
  {J.}~\bibnamefont {{Carretero}}}, \bibinfo {author} {\bibfnamefont
  {M.}~\bibnamefont {{Castellano}}}, \bibinfo {author} {\bibfnamefont
  {S.}~\bibnamefont {{Cavuoti}}}, \bibinfo {author} {\bibfnamefont
  {R.}~\bibnamefont {{Cledassou}}}, \bibinfo {author} {\bibfnamefont
  {G.}~\bibnamefont {{Congedo}}}, \bibinfo {author} {\bibfnamefont
  {L.}~\bibnamefont {{Conversi}}}, \bibinfo {author} {\bibfnamefont
  {L.}~\bibnamefont {{Corcione}}}, \bibinfo {author} {\bibfnamefont
  {M.}~\bibnamefont {{Cropper}}}, \bibinfo {author} {\bibfnamefont
  {F.}~\bibnamefont {{Dubath}}}, \bibinfo {author} {\bibfnamefont
  {S.}~\bibnamefont {{Dusini}}}, \bibinfo {author} {\bibfnamefont
  {G.}~\bibnamefont {{Fabbian}}}, \bibinfo {author} {\bibfnamefont
  {E.}~\bibnamefont {{Franceschi}}}, \bibinfo {author} {\bibfnamefont
  {M.}~\bibnamefont {{Fumana}}}, \bibinfo {author} {\bibfnamefont
  {B.}~\bibnamefont {{Garilli}}}, \bibinfo {author} {\bibfnamefont
  {F.}~\bibnamefont {{Grupp}}}, \bibinfo {author} {\bibfnamefont
  {H.}~\bibnamefont {{Hoekstra}}}, \bibinfo {author} {\bibfnamefont
  {F.}~\bibnamefont {{Hormuth}}}, \bibinfo {author} {\bibfnamefont
  {H.}~\bibnamefont {{Israel}}}, \bibinfo {author} {\bibfnamefont
  {K.}~\bibnamefont {{Jahnke}}}, \bibinfo {author} {\bibfnamefont
  {S.}~\bibnamefont {{Kermiche}}}, \bibinfo {author} {\bibfnamefont
  {B.}~\bibnamefont {{Kubik}}}, \bibinfo {author} {\bibfnamefont
  {M.}~\bibnamefont {{Kunz}}}, \bibinfo {author} {\bibfnamefont
  {F.}~\bibnamefont {{Lacasa}}}, \bibinfo {author} {\bibfnamefont
  {S.}~\bibnamefont {{Ligori}}}, \bibinfo {author} {\bibfnamefont {P.~B.}\
  \bibnamefont {{Lilje}}}, \bibinfo {author} {\bibfnamefont {I.}~\bibnamefont
  {{Lloro}}}, \bibinfo {author} {\bibfnamefont {E.}~\bibnamefont {{Maiorano}}},
  \bibinfo {author} {\bibfnamefont {O.}~\bibnamefont {{Marggraf}}}, \bibinfo
  {author} {\bibfnamefont {R.}~\bibnamefont {{Massey}}}, \bibinfo {author}
  {\bibfnamefont {S.}~\bibnamefont {{Mei}}}, \bibinfo {author} {\bibfnamefont
  {M.}~\bibnamefont {{Meneghetti}}}, \bibinfo {author} {\bibfnamefont
  {G.}~\bibnamefont {{Meylan}}}, \bibinfo {author} {\bibfnamefont
  {L.}~\bibnamefont {{Moscardini}}}, \bibinfo {author} {\bibfnamefont
  {C.}~\bibnamefont {{Padilla}}}, \bibinfo {author} {\bibfnamefont
  {S.}~\bibnamefont {{Paltani}}}, \bibinfo {author} {\bibfnamefont
  {F.}~\bibnamefont {{Pasian}}}, \bibinfo {author} {\bibfnamefont
  {S.}~\bibnamefont {{Pires}}}, \bibinfo {author} {\bibfnamefont
  {G.}~\bibnamefont {{Polenta}}}, \bibinfo {author} {\bibfnamefont
  {M.}~\bibnamefont {{Poncet}}}, \bibinfo {author} {\bibfnamefont
  {F.}~\bibnamefont {{Raison}}}, \bibinfo {author} {\bibfnamefont
  {J.}~\bibnamefont {{Rhodes}}}, \bibinfo {author} {\bibfnamefont
  {M.}~\bibnamefont {{Roncarelli}}}, \bibinfo {author} {\bibfnamefont
  {R.}~\bibnamefont {{Saglia}}}, \bibinfo {author} {\bibfnamefont
  {P.}~\bibnamefont {{Schneider}}}, \bibinfo {author} {\bibfnamefont
  {A.}~\bibnamefont {{Secroun}}}, \bibinfo {author} {\bibfnamefont
  {S.}~\bibnamefont {{Serrano}}}, \bibinfo {author} {\bibfnamefont
  {G.}~\bibnamefont {{Sirri}}}, \bibinfo {author} {\bibfnamefont {J.~L.}\
  \bibnamefont {{Starck}}}, \bibinfo {author} {\bibfnamefont {F.}~\bibnamefont
  {{Sureau}}}, \bibinfo {author} {\bibfnamefont {A.~N.}\ \bibnamefont
  {{Taylor}}}, \bibinfo {author} {\bibfnamefont {I.}~\bibnamefont {{Tereno}}},
  \bibinfo {author} {\bibfnamefont {R.}~\bibnamefont {{Toledo-Moreo}}},
  \bibinfo {author} {\bibfnamefont {L.}~\bibnamefont {{Valenziano}}}, \bibinfo
  {author} {\bibfnamefont {Y.}~\bibnamefont {{Wang}}},\ and\ \bibinfo {author}
  {\bibfnamefont {J.}~\bibnamefont {{Zoubian}}},\ }\bibfield  {title} {\bibinfo
  {title} {{Euclid: The reduced shear approximation and magnification bias for
  Stage IV cosmic shear experiments}},\ }\href
  {https://doi.org/10.1051/0004-6361/201937323} {\bibfield  {journal} {\bibinfo
   {journal} {Astronomy and Astrophysics}\ }\textbf {\bibinfo {volume} {636}},\
  \bibinfo {eid} {A95} (\bibinfo {year} {2020})},\ \Eprint
  {https://arxiv.org/abs/1912.07326} {arXiv:1912.07326 [astro-ph.CO]}
  \BibitemShut {NoStop}%
\bibitem [{\citenamefont {{Troxel}}\ and\ \citenamefont
  {{Ishak}}(2015)}]{2015PhR...558....1T}%
  \BibitemOpen
  \bibfield  {author} {\bibinfo {author} {\bibfnamefont {M.~A.}\ \bibnamefont
  {{Troxel}}}\ and\ \bibinfo {author} {\bibfnamefont {M.}~\bibnamefont
  {{Ishak}}},\ }\bibfield  {title} {\bibinfo {title} {{The intrinsic alignment
  of galaxies and its impact on weak gravitational lensing in an era of
  precision cosmology}},\ }\href
  {https://doi.org/10.1016/j.physrep.2014.11.001} {\bibfield  {journal}
  {\bibinfo  {journal} {Physics Report}\ }\textbf {\bibinfo {volume} {558}},\
  \bibinfo {pages} {1} (\bibinfo {year} {2015})},\ \Eprint
  {https://arxiv.org/abs/1407.6990} {arXiv:1407.6990 [astro-ph.CO]}
  \BibitemShut {NoStop}%
\bibitem [{\citenamefont {{Codis}}\ \emph {et~al.}(2015)\citenamefont
  {{Codis}}, \citenamefont {{Gavazzi}}, \citenamefont {{Dubois}}, \citenamefont
  {{Pichon}}, \citenamefont {{Benabed}}, \citenamefont {{Desjacques}},
  \citenamefont {{Pogosyan}}, \citenamefont {{Devriendt}},\ and\ \citenamefont
  {{Slyz}}}]{2015MNRAS.448.3391C}%
  \BibitemOpen
  \bibfield  {author} {\bibinfo {author} {\bibfnamefont {S.}~\bibnamefont
  {{Codis}}}, \bibinfo {author} {\bibfnamefont {R.}~\bibnamefont {{Gavazzi}}},
  \bibinfo {author} {\bibfnamefont {Y.}~\bibnamefont {{Dubois}}}, \bibinfo
  {author} {\bibfnamefont {C.}~\bibnamefont {{Pichon}}}, \bibinfo {author}
  {\bibfnamefont {K.}~\bibnamefont {{Benabed}}}, \bibinfo {author}
  {\bibfnamefont {V.}~\bibnamefont {{Desjacques}}}, \bibinfo {author}
  {\bibfnamefont {D.}~\bibnamefont {{Pogosyan}}}, \bibinfo {author}
  {\bibfnamefont {J.}~\bibnamefont {{Devriendt}}},\ and\ \bibinfo {author}
  {\bibfnamefont {A.}~\bibnamefont {{Slyz}}},\ }\bibfield  {title} {\bibinfo
  {title} {{Intrinsic alignment of simulated galaxies in the cosmic web:
  implications for weak lensing surveys}},\ }\href
  {https://doi.org/10.1093/mnras/stv231} {\bibfield  {journal} {\bibinfo
  {journal} {MNRAS}\ }\textbf {\bibinfo {volume} {448}},\ \bibinfo {pages}
  {3391} (\bibinfo {year} {2015})},\ \Eprint {https://arxiv.org/abs/1406.4668}
  {arXiv:1406.4668 [astro-ph.CO]} \BibitemShut {NoStop}%
\bibitem [{\citenamefont {{Asgari}}\ \emph {et~al.}(2021)\citenamefont
  {{Asgari}}, \citenamefont {{Lin}}, \citenamefont {{Joachimi}}, \citenamefont
  {{Giblin}}, \citenamefont {{Heymans}}, \citenamefont {{Hildebrandt}},
  \citenamefont {{Kannawadi}}, \citenamefont {{St{\"o}lzner}}, \citenamefont
  {{Tr{\"o}ster}}, \citenamefont {{van den Busch}}, \citenamefont {{Wright}},
  \citenamefont {{Bilicki}}, \citenamefont {{Blake}}, \citenamefont {{de
  Jong}}, \citenamefont {{Dvornik}}, \citenamefont {{Erben}}, \citenamefont
  {{Getman}}, \citenamefont {{Hoekstra}}, \citenamefont {{K{\"o}hlinger}},
  \citenamefont {{Kuijken}}, \citenamefont {{Miller}}, \citenamefont
  {{Radovich}}, \citenamefont {{Schneider}}, \citenamefont {{Shan}},\ and\
  \citenamefont {{Valentijn}}}]{2021A&A...645A.104A}%
  \BibitemOpen
  \bibfield  {author} {\bibinfo {author} {\bibfnamefont {M.}~\bibnamefont
  {{Asgari}}}, \bibinfo {author} {\bibfnamefont {C.-A.}\ \bibnamefont {{Lin}}},
  \bibinfo {author} {\bibfnamefont {B.}~\bibnamefont {{Joachimi}}}, \bibinfo
  {author} {\bibfnamefont {B.}~\bibnamefont {{Giblin}}}, \bibinfo {author}
  {\bibfnamefont {C.}~\bibnamefont {{Heymans}}}, \bibinfo {author}
  {\bibfnamefont {H.}~\bibnamefont {{Hildebrandt}}}, \bibinfo {author}
  {\bibfnamefont {A.}~\bibnamefont {{Kannawadi}}}, \bibinfo {author}
  {\bibfnamefont {B.}~\bibnamefont {{St{\"o}lzner}}}, \bibinfo {author}
  {\bibfnamefont {T.}~\bibnamefont {{Tr{\"o}ster}}}, \bibinfo {author}
  {\bibfnamefont {J.~L.}\ \bibnamefont {{van den Busch}}}, \bibinfo {author}
  {\bibfnamefont {A.~H.}\ \bibnamefont {{Wright}}}, \bibinfo {author}
  {\bibfnamefont {M.}~\bibnamefont {{Bilicki}}}, \bibinfo {author}
  {\bibfnamefont {C.}~\bibnamefont {{Blake}}}, \bibinfo {author} {\bibfnamefont
  {J.}~\bibnamefont {{de Jong}}}, \bibinfo {author} {\bibfnamefont
  {A.}~\bibnamefont {{Dvornik}}}, \bibinfo {author} {\bibfnamefont
  {T.}~\bibnamefont {{Erben}}}, \bibinfo {author} {\bibfnamefont
  {F.}~\bibnamefont {{Getman}}}, \bibinfo {author} {\bibfnamefont
  {H.}~\bibnamefont {{Hoekstra}}}, \bibinfo {author} {\bibfnamefont
  {F.}~\bibnamefont {{K{\"o}hlinger}}}, \bibinfo {author} {\bibfnamefont
  {K.}~\bibnamefont {{Kuijken}}}, \bibinfo {author} {\bibfnamefont
  {L.}~\bibnamefont {{Miller}}}, \bibinfo {author} {\bibfnamefont
  {M.}~\bibnamefont {{Radovich}}}, \bibinfo {author} {\bibfnamefont
  {P.}~\bibnamefont {{Schneider}}}, \bibinfo {author} {\bibfnamefont
  {H.}~\bibnamefont {{Shan}}},\ and\ \bibinfo {author} {\bibfnamefont
  {E.}~\bibnamefont {{Valentijn}}},\ }\bibfield  {title} {\bibinfo {title}
  {{KiDS-1000 cosmology: Cosmic shear constraints and comparison between two
  point statistics}},\ }\href {https://doi.org/10.1051/0004-6361/202039070}
  {\bibfield  {journal} {\bibinfo  {journal} {Astronomy \& Astrophysics}\
  }\textbf {\bibinfo {volume} {645}},\ \bibinfo {eid} {A104} (\bibinfo {year}
  {2021})},\ \Eprint {https://arxiv.org/abs/2007.15633} {arXiv:2007.15633
  [astro-ph.CO]} \BibitemShut {NoStop}%
\bibitem [{\citenamefont {{Harnois-D{\'e}raps}}\ \emph
  {et~al.}(2021)\citenamefont {{Harnois-D{\'e}raps}}, \citenamefont
  {{Martinet}}, \citenamefont {{Castro}}, \citenamefont {{Dolag}},
  \citenamefont {{Giblin}}, \citenamefont {{Heymans}}, \citenamefont
  {{Hildebrandt}},\ and\ \citenamefont {{Xia}}}]{2021MNRAS.506.1623H}%
  \BibitemOpen
  \bibfield  {author} {\bibinfo {author} {\bibfnamefont {J.}~\bibnamefont
  {{Harnois-D{\'e}raps}}}, \bibinfo {author} {\bibfnamefont {N.}~\bibnamefont
  {{Martinet}}}, \bibinfo {author} {\bibfnamefont {T.}~\bibnamefont
  {{Castro}}}, \bibinfo {author} {\bibfnamefont {K.}~\bibnamefont {{Dolag}}},
  \bibinfo {author} {\bibfnamefont {B.}~\bibnamefont {{Giblin}}}, \bibinfo
  {author} {\bibfnamefont {C.}~\bibnamefont {{Heymans}}}, \bibinfo {author}
  {\bibfnamefont {H.}~\bibnamefont {{Hildebrandt}}},\ and\ \bibinfo {author}
  {\bibfnamefont {Q.}~\bibnamefont {{Xia}}},\ }\bibfield  {title} {\bibinfo
  {title} {{Cosmic shear cosmology beyond two-point statistics: a combined peak
  count and correlation function analysis of DES-Y1}},\ }\href
  {https://doi.org/10.1093/mnras/stab1623} {\bibfield  {journal} {\bibinfo
  {journal} {MNRAS}\ }\textbf {\bibinfo {volume} {506}},\ \bibinfo {pages}
  {1623} (\bibinfo {year} {2021})},\ \Eprint {https://arxiv.org/abs/2012.02777}
  {arXiv:2012.02777 [astro-ph.CO]} \BibitemShut {NoStop}%
\bibitem [{\citenamefont {Sklar}(1959)}]{Skla59}%
  \BibitemOpen
  \bibfield  {author} {\bibinfo {author} {\bibfnamefont {A.}~\bibnamefont
  {Sklar}},\ }\bibfield  {title} {\bibinfo {title} {Fonctions de r\'epartition
  \`a n dimensions et leurs marges},\ }\href@noop {} {\bibfield  {journal}
  {\bibinfo  {journal} {Publications de l'Institut de Statistique de
  l'Universit\'e de Paris}\ }\textbf {\bibinfo {volume} {8}},\ \bibinfo {pages}
  {229} (\bibinfo {year} {1959})}\BibitemShut {NoStop}%
\end{thebibliography}%

\appendix

\section{Rewriting the trivariate PDF as functional of bivariate PDFs}
\label{app::B}

We aim here at expressing $\mP(\hkappa_1,\hkappa_2,\hkappa_3)$ purely as a function of $\mP(\hkappa_1,\hkappa_2)$ and $\mP(\hkappa_2,\hkappa_3)$. We know this is technically possible given equation~(\ref{eq::18}).

We start by writing the CGFs used as building blocks for the joint multivariate PDF in terms of univariate and bivariate PDFs using Laplace transformations as
\begin{align}
e^{\hphi(\lambda_2)}&=\int d\hkappa_2' \exp(\hkappa_2'\lambda_2) \mP(\hkappa_2')\,,\\
e^{\hphi_{\kappa_1}(\lambda_2)}&=\int d\hkappa_2' \exp(\hkappa_2'\lambda_2) \mP(\hkappa_1,\hkappa_2')\,,
\end{align}
and similarly for $\exp({\hphi_{\kappa_3}(\lambda_2)})$. We insert these into the expression for the trivariate PDF built from pairwise CGFs from equation~\eqref{P123} and thus finally obtain
\begin{align}
\label{eq::B3}
    &\mP(\hkappa_1,\hkappa_2,\hkappa_3)\\
 \notag   &=\int \frac{d\lambda_2}{2\pi \ii} e^{-\hkappa_2\lambda_2} 
    \frac{\int d\hkappa_2' e^{\hkappa_2'\lambda_2} [\mP(\hkappa_1,\cdot)*\mP(\cdot,\hkappa_3)](\hkappa_2')}{\int d\hkappa_2'' e^{\hkappa_2''\lambda_2} \mP(\hkappa_2'')}\,
\end{align}
where $\mP(\hkappa_1,\cdot)*\mP(\cdot,\hkappa_3)$ denotes the convolution of $\mP(\hkappa_1,\hkappa_2)$ and $\mP(\hkappa_2,\hkappa_3)$ w.r.t. $\hkappa_2$. This is a result of the convolution theorem stating that the product of the Laplace transforms of two functions can be written as the Laplace transform of their convolution. As expected from the somewhat complicated relationship between PDFs and CGFs, equation~(\ref{eq::B3}) is unfortunately not nearly as simple as equation~(\ref{eq::18}) though they both depict the exact same result without any loss of information.

\section{Joint PDFs in terms of copulas}
\label{app::copulas}

\subsection{Dissecting joint PDFs into marginals \& copulas}
According to Sklar's theorem \cite{Skla59}, any multivariate joint PDF $P(\{\hat\kappa_i\})$ can be written in terms of univariate marginal PDFs $P(\hat\kappa_i)$ and a copula which describes the dependence structure between the variables. Copulas are well-suited for high-dimensional statistical applications because they allow one to model and estimate joint probability distributions by estimating marginals and copulas separately. Copulas can be obtained from a range of parametric models that encode the dependence structure using just a few parameters and then compress the information of a high-dimensional multi-variate PDF down to a a set of uni- or bivariate marginal PDFs and a set of correlation parameters.

To fully disentangle the marginals from the correlation structure, one uses the values of the marginal cumulative distribution function (CDF) $u_i=C(\hat\kappa_i)\in [0,1]$ as variables. Those so-called filling factors $u_i$ are simply obtained from 
\begin{equation}
    u_i(\hat\kappa_i)=\mathcal C(\hat\kappa_i)=\int^{\hat\kappa_i}\!\!\! d{\hat\kappa_i}'\,  \mP({\hat\kappa_i}') \,.
\end{equation}
The marginal CDFs provide a ranking of variables in terms of cumulative probability and the filling factors $u_i$ allow to identify quantiles easily. The multivariate copula $C$ is now defined as the joint CDF $\mathcal C$ expressed in the new ranked variables $u_i$
\begin{equation}
    \label{eq:def_copula}
    C(\{u_i\})=\mathcal C(\{\hat\kappa_i=\mathcal C^{-1}(u_i)\}),
\end{equation}
Similarly as the PDF is obtained as derivative of the CDF, the copula density $c$ is obtained from a derivative of the copula 
\begin{align}
    c(u_1,\ldots,u_n)&=\frac{\partial C(u_1,\ldots,u_n)}{\partial u_1\cdots \partial u_n}\\
    &=\frac{\mP(\hat\kappa_1,\ldots,\hat\kappa_n)}{
\mP(\hat\kappa_1)\cdots \mP(\hat\kappa_n)} \Big|_{\hat\kappa_i=\mathcal C^{-1}(u_i)}\,,
\end{align}
which by the chain rule corresponds to the ratio of the joint PDF and the would-be joint PDF if the variables $\hat\kappa_i$ were  independent. The joint PDF can be reconstructed from the set of univariate marginals $\{\mP(\hat\kappa_i)\}$ and the multivariate copula. 

For a bivariate joint PDF, one can write 
\begin{equation}
\label{eq:def_copula_biv}
\mP(\hat\kappa_i,\hat\kappa_j)=\mP(\hat\kappa_i)\mP(\hat\kappa_j) c_{ij}(\mathcal C(\hat\kappa_i),\mathcal C(\hat\kappa_j))\,,
\end{equation}
where $c_{ij}$ is the copula density and if $\hat\kappa_i$ and $\hat\kappa_j$ are independent, then $c_{ij}=1$. There are many parametric copula families available to encode the dependence between $\mathcal C(\hat\kappa_i)$ and $\mathcal C(\hat\kappa_j)$ using a correlation parameter controlling the strength of dependence. A particularly simple case is the {\it Gaussian bivariate copula}, which is controlled by the pairwise correlation coefficient $r_{ij}$
\begin{equation}
\label{eq:def_copula_biv_Gauss}
\mP(\hat\kappa_i,\hat\kappa_j)=    \mP(\hat\kappa_i) \mP(\hat\kappa_j)c_{ij}^{\rm Gauss}(r_{ij})(\mathcal C(\hat\kappa_i),\mathcal C(\hat\kappa_j))\,.
\end{equation}

\begin{figure}
    \centering
    \includegraphics[width=\columnwidth]{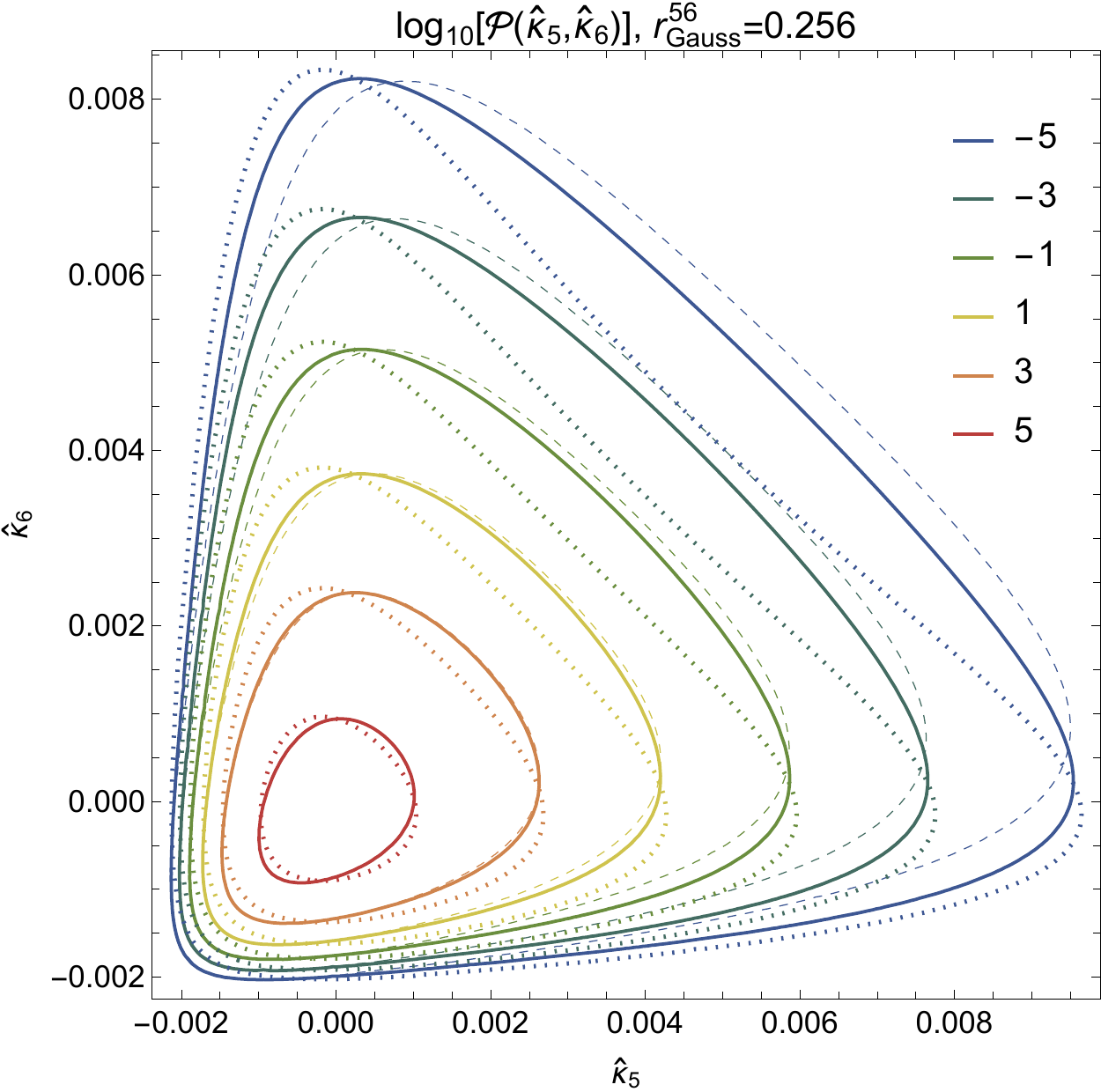}
    \caption{Contour plot of the joint PDF $P(\hat\kappa_5,\hat\kappa_6)$ between two neighbouring nulled convergence bins as determined by the LDT prediction for $P(\hat\kappa_5,\hat\kappa_6)$ (solid lines), the reconstructed PDF using the LDT prediction for the marginals $P(\hat\kappa_5)$ and $P(\hat\kappa_6)$ along with a Gaussian copula with correlation coefficient $r_{\rm Gauss}$ obtained from the LDT prediction (dashed lines) and the result one would obtain if the neighbouring $\hat\kappa$ bins were independent (dotted lines).}
    \label{fig:bivariatePDFcopula}
\end{figure}

We demonstrate a reconstruction of the joint PDF $\mP(\hat\kappa_5,\hat\kappa_6)$ using the marginals $\mP(\hat\kappa_{5/6})$ and a Gaussian copula density
in Figure~\ref{fig:bivariatePDFcopula}. It can be seen overall that the Gaussian copula approximation agrees very well with the exact LDT calculation in the bulk of the PDF and a bit less in the high and low convergence tails but with still a much better agreement than if we were to consider that the two $\hat{\kappa}$ bins were independent. A more rigorous assessment of the validity of the approximation would first require to assess the validity of the LDT/copula approaches with real or simulated data and then quantify the cosmological information content present in the bulk and tails of the PDFs, for example pursing a fisher analysis as in \cite{Boyle2020}.

\subsection{Measures of correlation}
The linear correlation coefficient is defined as usual as
\begin{align}
\label{eq:crosscorr}
r_{ij} = \text{Corr}(\hat\kappa_i,\hat\kappa_j) =\langle\hat\kappa_i\hat\kappa_j\rangle/\sqrt{\langle\hat\kappa_i^2\rangle\langle\hat\kappa_j^2\rangle} \,.
\end{align}
In our particular case, we note that the value of the correlation coefficients $r_{ij}$ will be closely related to the overlap between the nulled lensing kernels. For the case of discrete source planes, we found that the linear correlation coefficients measured from the joint bivariate PDFs are $r_{56}\simeq r_{67} \simeq 0.25$, which reflects the overlap between the neighbouring lensing kernels that are of the order of $25$\% in physical space quantified by the comoving distance rather than redshift as shown in Figure~\ref{fig:pofznum}.

For the modelling of copulas, one typically uses {\it ranked correlation coefficients} such as Spearman's-$\rho$. Spearman's-$\rho$ measures the correlation between the filling factors (obtained from the marginal CDFs),  
\begin{align}
\label{eq:Spearmancorr}
\rho_{S,ij}  &:= \text{Corr}(\mathcal C(\hat\kappa_i), \mathcal C(\hat\kappa_j)) \\
\notag &= 12\int_0^1\! du_i \int_0^1 \! d u_j\, C(u_i,u_j)-3
\,.
\end{align}
It can be thought of of the linear correlation coefficient of abundance-matched $\hat\kappa$-values. 
For a Gaussian bivariate copula, Spearman-$\rho$ is obtained as $\rho_S=\frac{6}{\pi} \arcsin(r/2)$ which is close to the linear correlation $r$. We invert this relation to obtain the Gaussian pairwise correlation coefficients $r_{ij}^{\rm Gauss}=2\sin(\pi\rho_{S,ij}/6)$ given the rank correlations $\rho_{S,ij}$. 

If one is more interested in the tails of the distribution, one would like to know whether with large/small values of $\hat\kappa_i$ also large/small values of $\hat\kappa_j$ are expected. This can be quantified using the tail concentration function defined for every value of cumulative probability $u$ as
\begin{align}
\label{eq:tailconcentration}
\lambda_C(u) = \begin{cases}
\frac{C(u,u)}{u} & u\in [0,0.5[\\
2 +\frac{C(u,u)-1}{1-u} & u\in[0.5,1]\,.
\end{cases}
\end{align}
The tail concentration measures the slope of the copula and upper or lower tail dependence is signalled if $\lambda_C(u\rightarrow 0)>0$ or $\lambda_C(u\rightarrow 1)>0$, respectively. 
\begin{figure}
    \centering
    \includegraphics[width=\columnwidth]{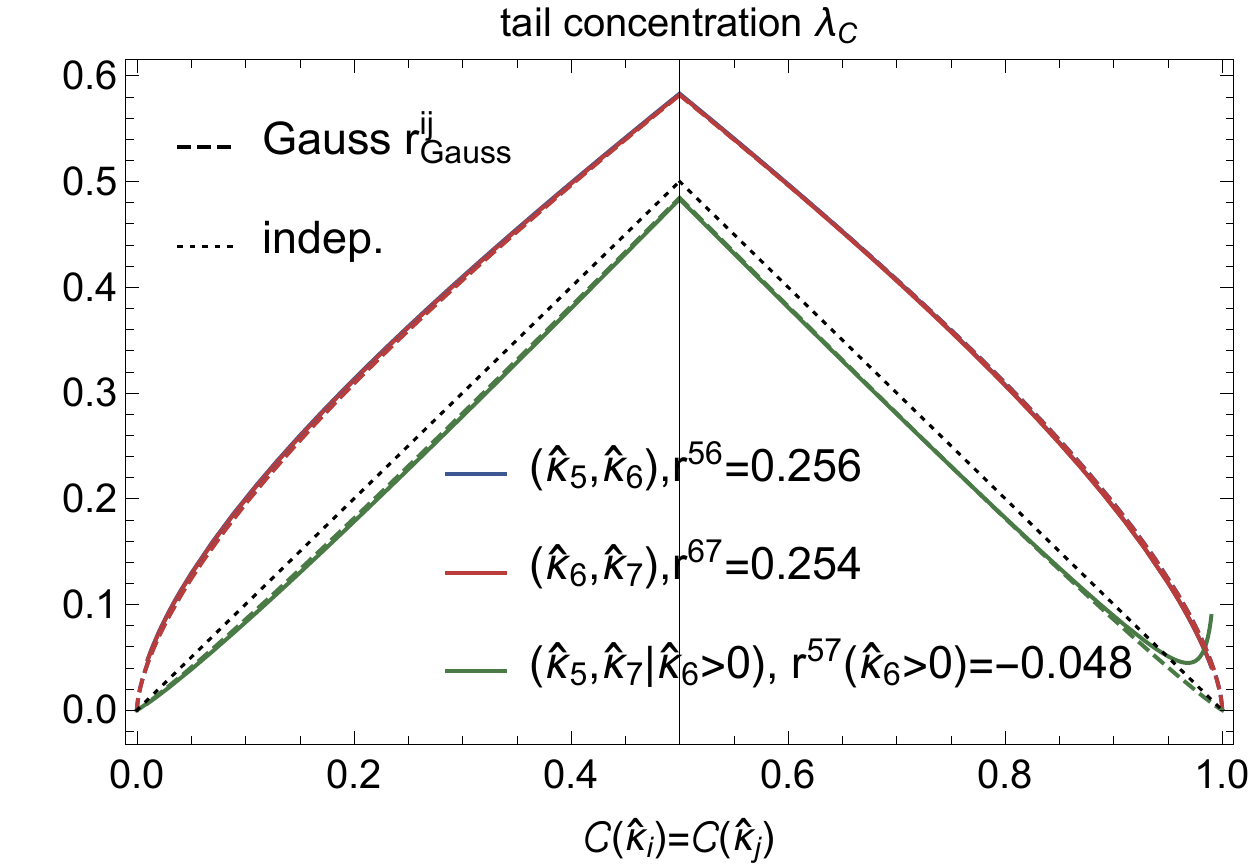}
    \caption{Tail correlation between nulled $\hat\kappa$ bins as predicted from LDT (solid lines) and reconstructed from Gaussian copulas with measured correlation (dashed lines) in comparison to the independent result (black dotted line). We show the almost identical results for two neighbouring bin pairs $\hat\kappa_{5/6}$ (blue) and $\hat\kappa_{6/7}$ (red) with virtually identical positive correlation and separated bins $\hat\kappa_{5/7}$ that are coupled through a constraint on the intermediate bin $\hkappa_6\gtrless 0$ leading to a slight negative correlation (green).}
    \label{fig:tailconcentration}
\end{figure}
We show the tail correlation for the joint PDF of neighbouring $\hkappa$ bins in Figure~\ref{fig:tailconcentration}. While they are positively correlated due to the overlap of their nulled lensing kernels, they do not exhibit upper or lower tail dependence, as $\lambda_C\rightarrow 0$ as one approaches the rare event tails $\mathcal C(\hkappa)\rightarrow 0$ and $\mathcal C(\hkappa)\rightarrow 1$, respectively. The correlation between regions of similar rarity is well described by a Gaussian copula (dashed lines), which does not exhibit any tail dependence. Since the $\hkappa_i$ have similar CDFs, the tail concentration is related to a diagonal slice of the joint PDF $\mP(\kappa_i,\hkappa_j)$ such as the one shown in Figure~\ref{fig:bivariatePDFcopula}.

Note that while a Gaussian bivariate copula does not exhibit any tail-dependence, conditionals obtained from a Gaussian multivariate copula or a Gaussian conditional copula can exhibit some tail dependence. We illustrate this point with the tail correlation between physically separated nulled bins $\hkappa_5$ and $\hkappa_7$ that is caused by a positivity constraint on the intermediate `coupling' bin $\hkappa_6>0$ as the green line in Figure~\ref{fig:tailconcentration}.

\begin{figure}
    \includegraphics[width=\columnwidth]{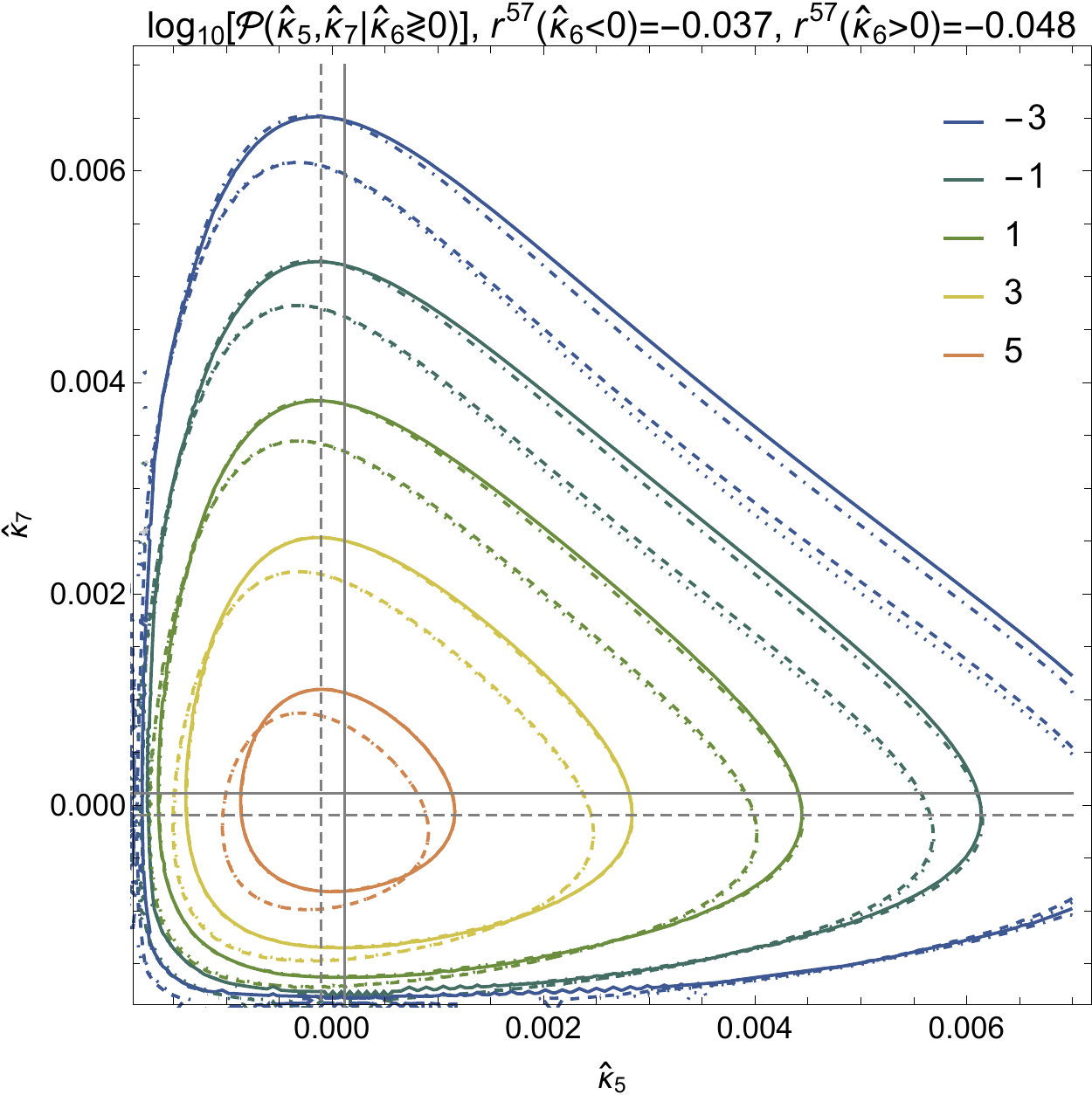}
    \caption{Contour plot of the joint conditional PDF $\mP(\hkappa_5,\hkappa_7|\hkappa_6\gtrless 0)$ between two nulled convergence bins coupled by a constraint on the intermediate bin $\hat\kappa_6\gtrless 0$ for the LDT prediction (solid/dashed lines) and the reconstructed PDF using the LDT prediction for the 
    bivariate marginals $\mP(\hkappa_5,\hkappa_6)$ and $\mP(\hkappa_6,\hkappa_7)$ along with a Gaussian conditional copula with correlation coefficients $r^{57}(\hkappa_6\gtrless 0)$ from LDT (dot-dashed/dotted lines). We also show the means of $\hkappa_{5/7}$ induced by $\hkappa>0$ (solid grey lines) and $\hkappa<0$ (dashed grey lines).}
    \label{fig:bivariateconditionalPDFcopula}
\end{figure}

\subsection{Modelling the trivariate PDF with a copula}

\subsubsection{Gaussian multivariate copula}
A Gaussian multivariate copula is controlled by a {\it correlation matrix} containing the pairwise correlations $r_{ij}$. For the case of discrete source planes, only neighbouring nulled $\hat\kappa$-bins are correlated. The simplest  starting point to extract information from an $n$-dimensional multivariate PDF would then be to rely on the set of $n$ univariate marginals $\{\mP(\hat\kappa_i)\}$ along with $n-1$ correlation coefficients. 

When comparing the full joint PDF to the `reconstruction' based on marginals and a multivariate Gaussian copula with the correlation coefficients as input, one can get an idea of how much information is lost. For example, the trivariate PDF $\mP(\hat\kappa_5,\hat\kappa_6,\hat\kappa_7)$ considered in the main text would be reconstructed using three univariate marginals and a trivariate Gaussian copula with correlation coefficients $r_{56}$, $r_{67}$ and $r_{57}=0$
\begin{align}
\label{eq:trivPDF_trivcopula}
 \mP(\hat\kappa_5,\hat\kappa_6,\hat\kappa_7)&=\mP(\hat\kappa_5)\mP(\hat\kappa_6)\mP(\hat\kappa_7)\\
 \notag &\times c_{567}^{\rm Gauss}(r_{56},r_{67})(\mathcal C(\hat\kappa_5),\mathcal C(\hat\kappa_6), \mathcal C(\hat\kappa_7))\,,
\end{align}
where the correlation coefficients $r_{56}$ and $r_{67}$ are determined from the joint bivariate PDFs as discussed in the context of equation~\eqref{eq:def_copula_biv_Gauss}.

\subsubsection{Conditional Gaussian bivariate copula}

The trivariate PDF $\mP(\hkappa_5,\hkappa_6,\hkappa_7)$ is constructed from the bivariate CGFs $\hphi(\lambda_5,\lambda_6)$ and $\hphi(\lambda_6,\lambda_7)$ and the univariate CGF $\hphi(\lambda_6)$. Since those CGFs can be constructed from the bivariate PDFs $\mP(\hkappa_5,\hkappa_6)$ and $\mP(\hkappa_6,\hkappa_7)$ along with the univariate PDF $\mP(\hkappa_6)$, it is natural to look for a reconstruction of the trivariate PDF in terms of bivariate PDFs and a conditional bivariate copula. 
\begin{align}
\label{eq:trivPDF_bivcopula}
 \mP(\hat\kappa_5,\hat\kappa_6,\hat\kappa_7)&=\frac{\mP(\hat\kappa_5,\hat\kappa_6)\mP(\hat\kappa_6,\hat\kappa_7)}{\mP(\hkappa_6)}\\
 \notag &\times c_{57|6}(\hkappa_6)\Big(\mathcal C(\hat\kappa_5|\hat\kappa_6), \mathcal C(\hat\kappa_7|\hat\kappa_6)\Big)\,.
\end{align}
For an illustration of this reconstruction, we will assume the conditional bivariate copula $c_{57|6}(\hkappa_6)=c_{57|6}^{\rm Gauss}(r_{57|6}(\hkappa_6))$ to be Gaussian and hence given by a correlation coefficient function $r_{57|6}(\hkappa_6)$ encoding the correlation coefficient between $\hkappa_5$ and $\hkappa_7$ given a value (or range) for $\hkappa_6$. While multivariate parametric copula models are only available for a few special classes such as Gaussian copulas, there is a plethora of bivariate parametric copula models that allows for more flexibility and could be used to refine the present model. We use the conditional Gaussian copula to reconstruct the bivariate conditional PDF $\mP(\hkappa_5,\hkappa_7|\hkappa_6\gtrless 0)$ where a coupling between $\hkappa_5$ and $\hkappa_7$ is induced by a constraint on $\hkappa_6\gtrless 0$, see Figure~\ref{fig:bivariateconditionalPDFcopula}. The good qualitative match of the Gaussian bivariate copula along a diagonal slice where $\mathcal C(\hkappa_5)=\mathcal C(\hkappa_7)$ can also be observed in the tail correlation shown in Figure~\ref{fig:tailconcentration}. Interestingly, we see that putting a positivity constraint on $\hkappa_6$ induces some positive tail dependence in the LDT prediction that is not captured by a simple Gaussian bivariate copula.
\end{document}